\documentclass[nofootinbib,twocolumn,prd,superscriptaddress,preprintnumbers,floatfix]{revtex4-1}
\usepackage{slashed,bm,graphicx,color,upgreek}
\usepackage{amssymb}
\usepackage{amsmath}
\usepackage{mathrsfs}
\usepackage[normalem]{ulem}
\RequirePackage[colorlinks=true,urlcolor=blue,anchorcolor=blue,citecolor=blue,filecolor=blue,
linkcolor=blue,menucolor=blue,linktocpage=true,pdfproducer=medialab]{hyperref}

\def\nn{\nonumber \\ }
\def\tr{ \text{Tr}\, }
\def\d{\mathsf{d}} 
\newcommand{\blue}[1]{\color{blue} #1 \color{black} }

\def\vpi{ {\bf \Pi} }
\def\xpi{ \pi }
\def\LL{\mathscr{L}}
\def\V{\mathbf{V}}
\def\T{\mathbf{T}}
\def\U{\mathbf{U}}
\def\cF{\mathcal{F}}
\def\cO{\mathcal{O}}

\def\vpiEW{ {\bf \Pi} }

\begin{document}

\title{Analysis of General Power Counting Rules in Effective Field Theory}%

\author{Belen Gavela}
\affiliation{\vspace{1mm} Instituto de F\'{\i}sica Te\'orica, IFT-UAM/CSIC,
Universidad Aut\'onoma de Madrid, Cantoblanco, 28049, Madrid, Spain}

\author{Elizabeth E.~Jenkins}
\affiliation{\vspace{1mm}
Department of Physics, University of California at San Diego, La Jolla, CA 92093, USA}
\affiliation{\vspace{1mm} CERN TH Division, CH-1211 Geneva 23, Switzerland}

\author{Aneesh V.~Manohar}
\affiliation{\vspace{1mm}
Department of Physics, University of California at San Diego, La Jolla, CA 92093, USA}
\affiliation{\vspace{1mm} CERN TH Division, CH-1211 Geneva 23, Switzerland}

\author{Luca Merlo}
\affiliation{\vspace{1mm} Instituto de F\'{\i}sica Te\'orica, IFT-UAM/CSIC,
Universidad Aut\'onoma de Madrid, Cantoblanco, 28049, Madrid, Spain}

\begin{abstract}
We derive the general counting rules for a quantum effective field theory (EFT) in $\d$ dimensions. The rules are valid for strongly and weakly coupled theories, and predict that all kinetic energy terms are canonically normalized. They determine the energy dependence of scattering cross sections in the range of validity of the EFT expansion.  We show that the size of cross sections is controlled by the $\Lambda$ power counting of EFT, not by chiral counting, even for chiral perturbation theory ($\chi$PT). The relation between $\Lambda$ and $f$ is generalized to $\d$ dimensions. We show that the naive dimensional analysis $4\pi$ counting is related to $\hbar$ counting. The EFT counting rules are applied to $\chi$PT, low-energy weak interactions, Standard Model EFT and the non-trivial case of Higgs EFT.
\end{abstract}

\date{\today}

\preprint{\blue{CERN-TH-2016-015}}
\preprint{\blue{FTUAM-16-2}}
\preprint{\blue{IFT-UAM/CSIC-16-006}}

\maketitle

%
%
\section{Introduction}

Effective field theories (EFT) have a wide application in high energy physics, and are often used to parametrize the (a priorit unknown) effects of new physics at low energies in terms of local operators. An EFT is a well-defined quantum field theory, and one can compute radiative corrections and renormalize the theory without  reference to any quantity outside the EFT. (See, e.g.\ Ref.~\cite{Manohar:1996cq} for a review and pedagogical examples.) 
EFT have a systematic expansion in a small parameter, usually called the EFT power counting formula, an example of which is Weinberg's momentum power counting formula for chiral perturbation theory ($\chi$PT)~\cite{Weinberg:1978kz}.  A power counting formula called naive dimensional analysis (NDA) applicable to general EFT has been known for some time~\cite{Manohar:1983md}. Recently, there has been renewed interest in EFT power counting and its application to new physics searches at the LHC~\cite{Jenkins:2013sda,Buchalla:2013eza,Buchalla:2014eca}. The use of EFT ideas, which have been well-established over several decades, to Higgs boson physics has caused considerable debate in the literature.

A generic effective Lagrangian describes many distinct fields and interactions, which naturally lead to independent expansions in different parameters such as momenta, couplings, etc.   In general, it is not possible to unify all expansions into a single expansion which is valid for all energy regimes.

In this paper, we start by presenting a pedagogical derivation of the EFT power counting rules for an arbitrary theory in $\d$ dimensions. The same power counting rules are valid for weakly and strongly interacting theories. We then show that the most general power counting rules are linear combinations of: 
\begin{enumerate}
\item[(a)] a counting rule for each coupling constant; 
\item[(b)] a counting rule for $\Lambda$, the EFT scale;  
\item[(c)] a counting rule for the total number of fields; 
\item[(d)] a generalization of Weinberg's momentum power counting rule which also counts fermion bilinears;
\item[(e)] a $4 \pi$ counting rule.
\end{enumerate}
The counting rules are not all linearly independent; and it is convenient to choose either (c) \emph{or} (d).   Rule (e) provides a good estimate for amplitudes and Lagrangian coefficients, and a convenient normalization which makes the distinction between strong and weak coupling regimes clear. One can reformulate (e) as an equivalent  $\hbar$ counting rule using the EFT loop expansion parameter $\hbar/(4\pi)^{\d/2}$, as explained in Sec.~\ref{Sect:MasterFormula}.

The NDA power counting rule is the usual $\Lambda$ counting rule for EFT combined with the  $4\pi$ counting rule, which provides a systematic way to normalize EFT operators~\cite{Manohar:1983md}. One is free to pick any operator normalization by redefining the operator coefficients; the advantages of using NDA normalization are discussed in Section~\ref{sec:eg}.

The power counting formula in the form given here leads to the correct normalization for terms in the Standard Model (SM) Lagrangian, including the kinetic terms, and it gives a homogeneous counting rule for the two terms in the gauge covariant derivative $D=\partial + i g A$, where $g$ and $A$ denote a generic gauge coupling and gauge boson, respectively. We also find an interesting generalization of the relation $\Lambda \sim 4 \pi f$ in $\d=4$ dimensions to $\d$ dimensions.

We provide many examples of the use of power counting rules in Sec.~\ref{sec:eg}. In particular, we discuss the well-known applications to $\chi$PT and the Fermi theory of low-energy weak interactions. The Fermi theory is an instructive example of how low-energy measurements, which can be computed using the EFT, can be used to determine the pattern of operator coefficients and hence the structure of the underlying UV theory. The Fermi theory is subtle --- the power counting rules depend on the operator flavor structure. 
 The Fermi theory also demonstrates that  ``one scale, one coupling" power counting~\cite{Panico:2015jxa} is not valid for the low-energy limit of spontaneously broken gauge theories.  We  discuss the application of the counting rules to the Standard Model EFT (SMEFT) and how the renormalization group evolution obeys  NDA normalization. We give a toy example that illustrates the utility of NDA counting in an explicit matching computation. As a final example, we explain the counting rule for triple-gauge operators and anomalous magnetic moments.

We apply the counting rules to study scattering cross sections in Sec.~\ref{sec:compare}, and show that the size of cross sections is given by the usual $\Lambda$ counting rule. This result applies not only to the SMEFT -- also known as linear electroweak EFT -- but also to other effective theories such as $\chi$PT and non-linear Higgs electroweak EFT (HEFT). 

Application of the power counting rules to HEFT is discussed in Sec.~\ref{sec:heft}.  Conclusions are presented in Sec.~\ref{sec:conclusions}.
%
%
\section{Master Formula in $\d$ Dimensions}
\label{Sect:MasterFormula}

This section presents a pedagogical discussion of EFT power counting rules. Many of the results are known. This section establishes notation,  makes clear the connection between different counting rules, and   summarizes results which will be used in the later discussion. The results are given in $\d$ dimensions; the extension of the $\d=4$ results to arbitrary dimension is new, and the $\d$ dependence of the relation between $\Lambda$ and $f$ is non-trivial.

In $\d$ spacetime dimensions, the mass dimensions of generic scalar $\phi$, gauge boson $A$ and fermion $\psi$ fields and of generic gauge $g$, Yukawa $y$ and quartic scalar $\lambda$ coupling constants are
\begin{align}
\begin{aligned}
{}[\phi] &= \frac{\d-2}{2}, & {}[A] &= \frac{\d-2}{2}, &
{}[\psi] &= \frac{\d-1}{2}, \\
{}[g] &= \frac{4-\d}{2}, &
{}[y] &= \frac{4-\d}{2}, &
{}[\lambda] &= 4-\d .
\end{aligned}
\label{1}
\end{align}
The couplings $g$, $y$ and $\lambda$ denote generic gauge, Yukawa and quartic scalar coupling constants in the dimension $\le \d$ terms of the EFT Lagrangian.
Cubic scalar couplings, and scalar and fermion mass terms also can be included, as in Ref.~\cite{Jenkins:2013sda}.  The generalization to include these additional couplings is discussed below. 
The scalar fields $\phi$ include both Goldstone boson and non-Goldstone boson fields.  We will assume that the kinetic terms in the Lagrangian are canonically normalized, and that a general interaction term has the form
\begin{align}
\partial^{N_{p,i}} \phi^{N_{\phi,i}}  A^{N_{A,i}} \psi^{N_{\psi,i}} \Lambda^{N_{\Lambda,i}} (4\pi)^{N_{4\pi,i}}g^{N_{g,i}} y^{N_{y,i}} \lambda^{N_{\lambda,i}} ,
\label{2}
\end{align}
where $N_{a,i}$ counts the number of factors of each type $a$ appearing in the vertex $i$, and we have included factors of $4 \pi$ in the normalization to allow us to count $4\pi$ factors arising from loops. Since higher dimension operators are suppressed by inverse powers of $\Lambda$, $N_\Lambda < 0$ for operators with dimension $\ge \d$.

The powers in Eq.~(\ref{2}) are not all independent, but satisfy the constraint that terms in the Lagrangian have mass dimension $\d$ in $\d$ spacetime dimensions,
\begin{equation}
\begin{split}
\d = N_{p,i} + \frac{\d-2}{2} (N_{\phi,i}+N_{A,i})+ \frac{\d-1}{2} N_{\psi,i}+ N_{\Lambda,i }   \\
 + \frac{4-\d}{2} (N_{g,i}+N_{y,i}+2N_{\lambda,i})\,.
\end{split}
\label{11}
\end{equation}

If we define the total number of fields at a vertex by
\begin{align}
N_{F,i} \equiv N_{\phi,i}+N_{A,i}+N_{\psi,i},
\label{11a}
\end{align}
then Eq.~(\ref{11}) can be written as
\begin{equation}
\begin{split}
0 =\left( N_{p,i}+ \frac{1}{2} N_{\psi,i}-2 \right) + \frac{\d-2}{2} (N_{F,i}-2) + N_{\Lambda,i }\\
+ \frac{4-\d}{2} (N_{g,i}+N_{y,i}+2N_{\lambda,i}),
\end{split}
\label{11c}
\end{equation}
which will be useful later.

Now consider an arbitrary connected EFT diagram with insertions of vertices of the type Eq.~(\ref{2}). The diagram will generate an amplitude of the same form as in Eq.~(\ref{2}), and we can determine the powers of the different factors.

The number of external fields are given by
\begin{subequations}
\begin{align}
\label{3a}
N_\phi &= \sum_i N_{\phi,i} - 2 I_\phi,\\
\label{3b}
N_\psi &= \sum_i N_{\psi,i} - 2 I_\psi,\\
N_A &= \sum_i N_{A,i} - 2 I_A ,
\label{3c}
\end{align}
\end{subequations}
where the sum is over all vertices, and $I_{\phi,\psi,A}$ are the number of internal lines of each type, since each internal line results from the contraction of two fields. 

The dependence of the overall diagram on each coupling constant is
\begin{subequations}
\begin{align}
\label{g}
N_{g} &= \sum_i N_{g,i} \,, \qquad (R_g)\\
\label{y}
N_{y} &= \sum_i N_{y,i} \,,  \qquad (R_y)\\
\label{lambda}
N_{\lambda} &= \sum_i N_{\lambda,i} \,,  \qquad (R_\lambda)
\end{align}
\end{subequations}
where $R_{g,y,\lambda}$ denote these relations. These equations are obvious, because coupling constants appear only in the Feynman rules of vertices, since all propagators have been normalized to unity and do not bring in factors of the coupling, nor do the loop integrals.

The $\Lambda$ dependence of the amplitude is
\begin{align}
N_\Lambda &= \sum_i N_{\Lambda,i} \,. \qquad (R_\Lambda)
\label{Lambda}
\end{align}
This equation follows trivially because loop integrals do not generate powers of $\Lambda$ if one uses a mass-independent regulator such as dimensional regularization.

The $4\pi$ factors are given by
\begin{align}
N_{4\pi} &= \sum_i N_{4\pi,i} - \frac {\d}{2} L , \qquad (R_{4\pi})
\label{4pi}
\end{align}
where $L$ is the number of loops, since each loop in $\d$ dimensions has an overall factor of $(4\pi)^{-\d/2}$. 

The overall power of momentum $p$ is
\begin{align}
N_{p} &= \sum_i N_{p,i} -2 I_\phi - 2 I_A - I_\psi + \d L
\label{p}
\end{align}
since each internal scalar or gauge propagator is order $1/p^2$, each fermion propagator is order $1/p$, and a loop integral is order $p^\d$. For $p$ counting, light particle masses are treated as mass insertions. Otherwise, Eq.~(\ref{p}) gives the maximum power of $p$, and one can get lower powers with some $p \to m$, or equivalently, one has to count the sum of powers of $p$ and $m$, supplementing Eq.~(\ref{2}) by an additional $m$-counting factor.

In addition, there is one graph theory identity
\begin{align}
V - I + L &=1,
\label{euler}
\end{align}
which holds for a \emph{connected} graph. Here $V$ is the total number of vertices,  $I=I_\phi+I_\psi+I_A$ is the total number of internal lines, and $L$ is the number of loops, and the r.h.s.\ is the Euler character $\chi=1$ for a connected graph.

In summary, we have 10 relations Eq.~(\ref{3a})--(\ref{euler}).  Using Eqs.~(\ref{3a})--(\ref{3c}) to eliminate $I_{\phi,\psi,A}$ leaves 7 relations that do not depend on the number of internal lines. Five relations are $R_{g,y,\lambda,\Lambda,4\pi}$.  Eliminating $I$ in Eqs.~(\ref{3a})--(\ref{3c}) and~(\ref{euler}) gives the field counting rule 
\begin{align}
N_{F}-2 &= \sum_i \left(N_{F,i}-2\right)-2L , \qquad (R_F)
\label{field}
\end{align}
where $N_{F,i}$ is defined in Eq.~(\ref{11a}). 
The final relation, obtained from Eqs.~(\ref{p}) and~(\ref{euler}), is
\begin{align}
N_{\chi} -2 &= \sum_i \left( N_{\chi,i}-2\right) + (\d-2) L , \quad (R_\chi)
\label{chi}
\end{align}
where
\begin{eqnarray}
N_{\chi} \equiv N_{p} + \frac{N_{\psi}}{2}\,, \quad\quad
N_{\chi,i} \equiv N_{p,i} + \frac{N_{\psi,i}}{2} \,.
\label{19}
\end{eqnarray}
Eq.~(\ref{chi}) is a generalization of Weinberg's power counting formula for $\chi$PT~\cite{Weinberg:1978kz} to $N_\psi \not=0$.  It also gives the power counting rule for baryon $\chi$PT~\cite{Jenkins:1990jv,Jenkins:1991es} with $N_\psi=2$. $N_\chi$ will be used extensively in the $\chi$PT discussion later in this article. $N_\chi$ is related to the chiral dimension defined in  Refs.~\cite{Buchalla:2013eza,Buchalla:2014eca}, and discussed after Eq.~(\ref{36}).

The linear combination
\begin{equation}
\begin{split}
\frac{\d-2}{2} (N_F-2) +& (N_\chi-2) + N_\Lambda +\\
&+\frac{4-\d}{2} (N_g+N_y+2N_\lambda)=0\,,
\end{split}
\label{reln2}
\end{equation}
is equivalent to Eq.~(\ref{11}), i.e.\ the Lagrangian term induced by loop graphs also has mass dimension $\d$ if all the interaction Lagrangian terms have mass dimension $\d$. Only 6 of the 7 relations are linearly independent because of this constraint, which becomes
\begin{align}
(N_F-2) +& (N_\chi-2) + N_\Lambda =0\,,
\label{reln2a}
\end{align}
in $\d=4$.

Of the 7 relations $R_i$, $i=g,y,\lambda,\Lambda,4\pi,F,\chi$, the first four are independent of the number of loops $L$ in the EFT graph.  However, by replacing $R_{4\pi}$ by $R^\prime_{4\pi} \equiv R_{4\pi}-(\d/4)R_F$,
\begin{align}
N_{4\pi}-\frac{\d}{4}\left( N_F - 2 \right) &= \sum_i \left( N_{4\pi,i}-\frac{\d}{4}N_{F,i}+\frac{\d}{2} \right), \qquad (R^\prime_{4\pi})
\label{4piprime}
\end{align}
and $R_\chi$ by $R_\chi^\prime \equiv R_\chi +[(\d -2)/2] R_F$,
\begin{align}
\left( N^\prime_\chi - \d \right) 
&= \sum_i   \left( N^\prime_{\chi,i} -\d \right), \qquad (R^\prime_{\chi})
\label{7b}
\end{align}
where
\begin{align}
N^\prime_{\chi,i} &\equiv N_{p,i} + \frac{N_{\psi,i}}{2} + \frac{\d-2}{2} N_{F,i} ,
\label{chiprime}
\end{align}
one obtains two alternative relations which are independent of the number of loops. Thus, of the 7 relations $R_g$, $R_y$, $R_\lambda$, $R_\Lambda$, $R_F$, $R_{4 \pi}^\prime$ and $R^\prime_\chi$, only one relation $R_F$ depends on the number of loops.  As before, only 6 of the 7 relations are independent because of the constraint Eq.~(\ref{11}).

Any $4\pi$ counting rule that is independent of the number of loops must satisfy $R^\prime_{4\pi}$. For the rule to be self-consistent, one must obtain the same $4\pi$ counting for an operator independently of the way it is generated. This  implies that the individual terms of Eq.~(\ref{4piprime}) must vanish, or be a linear combination of the other independent conserved quantities, namely the couplings and $\Lambda$.  Requiring that they vanish implies
\begin{align}
N_{4\pi} &= \frac{\d}{4}\left( N_F- 2 \right) \,,
\label{4piscaling}
\end{align}
which is tantamount to the statement that each field scales with a factor of $(4\pi)^{\d/4}$ and that the overall Lagrangian scales with  $(4 \pi)^{-\d/2}$. The more general case where individual terms in the sum in Eq.~(\ref{4piprime}) are a linear combination of conserved quantities is equivalent to the additional freedom to rescale couplings and $\Lambda$ by factors of $4\pi$ while still satisfying Eq.~(\ref{4piprime}). We can take advantage of this freedom to use, instead of Eq.~(\ref{4piscaling}),
\begin{align}
N_{4\pi} &= \frac{\d}{4}\left( N_F- 2 \right) -\frac{\d}{4} \left( N_g + N_y + 2 N_\lambda \right) \,,
\label{4piNDAscaling}
\end{align}
which is the $4\pi$ scaling rule of NDA~\cite{Manohar:1983md} in $\d$ dimensions. This choice of scaling gives canonical gauge boson kinetic terms.
Combined with the usual EFT power counting in $\Lambda$ dictated by dimensional analysis, Eq.~(\ref{4piNDAscaling}) gives the NDA master formula. 

The NDA master formula in $\d$ dimensions is that each operator in the Lagrangian is normalized according to
\begin{widetext}
\begin{align}
&\frac{\Lambda^\d}{(4\pi)^{\d/2} } \! \left[\frac{\partial}{\Lambda}\right]^{N_p} \! \left[\frac{( 4 \pi)^{\d/4}  \phi}{ \Lambda^{(\d-2)/2}} \right]^{N_\phi}\! \left[\frac{( 4 \pi)^{\d/4} A}{ \Lambda^{(\d-2)/2}}\right]^{N_A}\! \left[\frac{( 4 \pi)^{\d/4}  \psi}{\Lambda^{(\d-1)/2}}\right]^{N_\psi} \! \left[\frac{g}{(4 \pi)^{\d/4} \Lambda^{(4-\d)/2} }  \right]^{N_g}\!
\left[\frac{y}{(4 \pi)^{\d/4} \Lambda^{(4-\d)/2} } \right]^{N_y}\! \left[\frac{\lambda}{(4 \pi)^{\d/2} \Lambda^{(4-\d)} }\right]^{N_\lambda}\!  
\label{nda}
\end{align}
times coefficients of order unity.  In $\d=4$, this formula reduces to
\begin{align}
&\frac{\Lambda^4}{16 \pi^2 } \left[\frac{\partial}{\Lambda}\right]^{N_p}  \left[\frac{ 4 \pi\,  \phi}{ \Lambda} \right]^{N_\phi}
 \left[\frac{ 4 \pi\,  A}{ \Lambda } \right]^{N_A}  \left[\frac{ 4 \pi \,  \psi}{\Lambda^{3/2}}\right]^{N_\psi} \left[ \frac{g}{4 \pi }  \right]^{N_g}
\left[\frac{y}{4 \pi } \right]^{N_y} \left[\frac{\lambda}{16 \pi^2 }\right]^{N_\lambda}  ,
\label{nda4}
\end{align}
which is the NDA rule in the form given in Ref.~\cite{Cohen:1997rt,Luty:1997fk}. 
\end{widetext}

From the master formula Eq.~(\ref{nda}), it follows that the kinetic energy terms are of order
\begin{align}
i \overline \psi \slashed{\partial} \psi, \qquad (\partial_\mu \phi)^2, \qquad {X}_{\mu \nu}^2\,,
\label{kin}
\end{align}
where $X_{\mu\nu}$ is a generic field strength tensor, and the gauge, Yukawa and scalar interaction terms in $\d=4$ become
\begin{align}
g \overline \psi \slashed{A} \psi,\quad y \overline \psi \psi \phi,\quad \lambda \phi^4
\label{15}
\end{align}
times factors of order unity, which are  the conventional normalizations. 

The NDA rule Eq.~(\ref{nda}) also works if the coupling constant is absorbed into the gauge field, $A^\prime=g A$. In this case, the counting rule is $A^\prime/\Lambda$ for the new gauge field, and the kinetic term is normalized to
\begin{align}
\frac{1}{g^2} X_{\mu \nu}X^{\mu \nu}
\label{kin2}
\end{align}
times a coefficient of order unity. 

Finally, a generic four-fermion operator arises with a suppression by two powers of the EFT scale $\Lambda$, 
\begin{equation}
\dfrac{(4\pi)^2}{\Lambda^2}\overline \psi \psi \overline \psi \psi\,,
\label{generic4f}
\end{equation}
independent of the Lorentz contraction performed to construct the operator and with no assumptions about how the operator originates from integrating out heavier particles.

The power counting rule in Eqs.~(\ref{nda},\ref{nda4}) is actually an inequality~\cite{Manohar:1983md}. The reason is that certain operators can have small coefficients, and this does not affect the overall power counting. However, one cannot have large coefficients, because otherwise loop graphs involving these operators will generate large coefficients in other operators as well. A trivial example is dimension-six operators in the Fermi theory of weak interaction. One can have $\Delta B=1$ baryon number violating operators with coefficients much smaller than $G_F$, and this is consistent with the power counting. However, it is not possible to have such operators with coefficients much larger than $G_F$ --- other than the phenomenological problem of instantaneous proton decay, such operators would produce $\Delta B=0$ operators with coefficients much larger than $G_F$ through graphs involving $\Delta B=1$ and $\Delta B=-1$ operators.

A quick way of understanding the $4\pi$ factors in the NDA master formula is to recall that $\hbar$ counts the number of loops in the EFT  (using the action $S/\hbar$),  so that the EFT loop expansion is in powers of $\hbar/(4\pi)^{\d/2}$. Setting $\hbar=(4\pi)^{\d/2}$, so that $\hbar$ cancels the loop factor, and noting that quantum fields have dimension $\hbar^{1/2}$ gives Eq.~(\ref{nda}).  The $4\pi$ redefinitions of coupling constants discussed in Sec.~\ref{sec:eg} are given by their $\hbar$ dimensions. $4\pi$ counting is equivalent to $\hbar$ counting in the EFT,  and this formulation gives an equivalent version of counting rule (e). $\hbar$ counting has also been discussed previously in Ref.~\cite{Elias-Miro:2013mua,espinosa}. 

In the derivation of Eq.~(\ref{nda}), the number of loops refers to graphs in the EFT. It is not possible, in general, to count loops in the UV theory and to assign a loop order to couplings in the EFT. An example is $\chi$PT, where the scalar pion sector arises from strong dynamics. In this case, the low-energy degrees of freedom are non-perturbative, and $\hbar$ counting of the UV theory does not survive in the EFT. The low-energy dynamics is governed by $f \propto \Lambda_{\text{QCD}}$, and
\begin{align}
\left( \frac{\Lambda_{\text{QCD}}}{\mu} \right)^{b_0} &= e^{-8\pi^2/\left[\hbar g_3^2(\mu)\right]}
\label{24}x
\end{align}
where $g_3(\mu)$ is the QCD gauge coupling, and $b_0$ is the first term in the QCD $\beta$-function. Eq.~(\ref{24}) is non-analytic in $\hbar$ and $g_3$, and one cannot assign a QCD loop order to terms in the chiral Lagrangian. A more detailed discussion of this point, as well as related aspects of minimal coupling are discussed in detail in Ref.~\cite{Jenkins:2013fya}.

Note that \emph{defining} $\Lambda$ to be the scale of the momentum expansion in powers of $p/\Lambda$, as in Eqs.~(\ref{nda}) and (\ref{nda4}),  eliminates the freedom to rescale $\Lambda$ by powers of $4\pi$.
The advantage of rescaling the gauge coupling according to NDA Eq.~(\ref{4piNDAscaling}), rather than the simpler choice Eq.~(\ref{4piscaling}), is that the gauge covariant derivative
\begin{align}
\frac{D}{\Lambda} &=  \frac{\partial}{\Lambda} +i\left[ \frac{g}{4 \pi^{\d/4} \Lambda^{(4-\d)/2} }  \right] \left[ \frac{( 4 \pi)^{\d/4}  A}{ \Lambda^{(\d-2)/2}}\right]
= \frac{\partial + i g A}{\Lambda}
\label{16}
\end{align}
has a homogeneous power counting, since the $4\pi$ scalings of $g$ and $A$ cancel in the product.  The gauge boson field scales in the same manner as the scalar field.  Note that the covariant derivative does \emph{not} have homogeneous scaling in $N_\chi$; $\partial$ has $N_\chi=1$ and $gA$ has $N_\chi=0$.

It is straightforward to extend Eqs.~(\ref{nda}) and~(\ref{nda4}) to include scalar mass terms $m_{\phi}^2 \phi^2$, fermion mass terms $m_{\psi} \overline \psi \psi$, and trilinear scalar couplings $\kappa \phi^3$~\cite{Jenkins:2013sda}:
\begin{equation}
\left[\frac{m_{\phi}^2}{\Lambda^2}\right]^{N_{m_{\phi}}} \left[\frac{m_{\psi}}{\Lambda}\right]^{N_{m_{\psi}}} \left[\frac{\kappa}{(4 \pi)^{\d/4} \Lambda^{(6-\d)/2} } \right]^{N_\kappa}\,,
\end{equation}
which in $\d=4$ reduces to 
\begin{equation}
\left[\frac{m_{\phi}^2}{\Lambda^2}\right]^{N_{m_{\phi}}} \left[\frac{m_{\psi}}{\Lambda}\right]^{N_{m_{\psi}}} \left[\frac{\kappa}{4 \pi \Lambda } \right]^{N_\kappa}\,.
\end{equation}

In $\chi$PT in $\d=4$, the Lagrangian is a function of $U \equiv \exp2 i \vpi/f$, where $\vpi(x)$ is the pion matrix, and $f$ is the pion decay constant. In $\d$ dimensions,  $U=\exp 2i \vpi/f^{(d-2)/2}$, since $f$ has mass dimension one.  Expanding $U$ gives arbitrary powers of $\vpi/f^{(\d-2)/2}$. Comparing with Eq.~(\ref{nda}), we see that $f^{(\d-2)/2}$ must be the denominator for scalar fields. This result fixes the relation between $\Lambda$ and $f$,\footnote{Two dimensions is special~\cite{Coleman:1973ci}.}
\begin{align}
\Lambda &= (4\pi)^{\frac{\d}{2(\d-2)}}f, & \d &\not=2 .
\label{17}
\end{align}
\begin{figure}
\includegraphics[width=8cm]{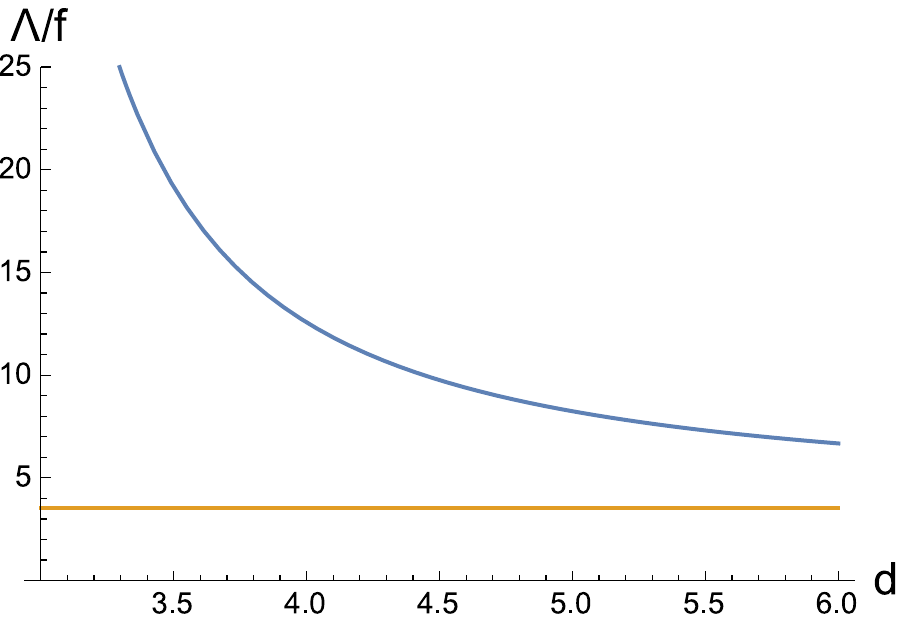}
\caption{\label{lplot} \it Plot of $\Lambda/f$ as a function of spacetime dimension $\d$. The asymptotic value as $d \to \infty$ is $\sqrt{4\pi}$, shown as the horizontal line.}
\end{figure}
The ratio $\left( \Lambda/f \right)$ has an interesting dependence on $\d$, which is shown in Fig.~\ref{lplot}.  Strictly speaking, as shown in Ref.~\cite{Manohar:1983md}, Eq.~(\ref{17}) is an inequality $\Lambda \leqslant (4\pi)^{\frac{\d}{2(\d-2)}}f$ rather than an equality.
The NDA master formula Eq.~(\ref{nda}) written using both $\Lambda$ and $f$ becomes
\begin{widetext}
\begin{align}
&f^{d-2} \Lambda^{2} \left[\frac{\partial}{\Lambda}\right]^{N_p} \left[\frac{\phi}{ f^{(\d-2)/2} } \right]^{N_\phi} \left[\frac{A}{ f^{(\d-2)/2} } \right]^{N_A} \left[\frac{\psi}{ f^{(\d-2)/2} \sqrt{\Lambda}} \right]^{N_\psi}  \left[\frac{g}{(4 \pi)^{\d/4} \Lambda^{(4-\d)/2} }  \right]^{N_g}\!
\left[\frac{y}{(4 \pi)^{\d/4} \Lambda^{(4-\d)/2} } \right]^{N_y}\! \left[\frac{\lambda}{(4 \pi)^{\d/2} \Lambda^{(4-\d)} }\right]^{N_\lambda}\!  
\label{fnda}
\end{align}
which reduces in $\d=4$ to~\cite{Manohar:1983md}
\begin{align}
&f^{2} \Lambda^{2} \left[\frac{\partial}{\Lambda}\right]^{N_p} \left[\frac{\phi}{ f } \right]^{N_\phi} \left[\frac{A}{ f } \right]^{N_A} \left[\frac{\psi}{ f \sqrt{\Lambda}} \right]^{N_\psi}  \left[\frac{g}{ 4\pi }  \right]^{N_g}
 \left[\frac{y}{ 4\pi } \right]^{N_y} \left[\frac{ \lambda}{ 16\pi^2}\right]^{N_\lambda}.
\label{fnda4}
\end{align}
\end{widetext}

Example applications of Eq.~(\ref{nda4}) are given in Sec.~\ref{sec:eg}. The factors of $4\pi$ in NDA are of practical importance. For example, in $\chi$PT, the derivative expansion is in powers of $p/\Lambda$, and is valid for pion scattering with momenta smaller than $\Lambda = 4 \pi f \sim 1$\, GeV, where $f$ is the pion decay constant.  Since $m_\pi \ge f$, using $f$, instead of $\Lambda$, as the momentum expansion parameter would indicate that $\chi$PT should fail even for $\xpi\xpi$ scattering at threshold, where $p \sim m_\pi$, which is not the case.

In summary, the \emph{only} counting rules that are consistent are combinations of:
\begin{itemize}
\item[--] The coupling constant rules $R_{g,y,\lambda}$.
\item[--] $R_\Lambda$, which is the usual EFT rule counting powers of $\Lambda$.  Since in EFTs, $N_\Lambda<0$, this rule counts powers of $\Lambda$ in the \emph{denominator}. 
\item[--] The $4\pi$ counting rule $R_{4\pi}^\prime$ (or equivalently $R_{4\pi}$).
\item[--] The field counting rule $R_F$.
\item[--] The momentum rule $R_\chi$ (or equivalently $R^\prime_\chi$), which counts $N_\chi \equiv N_p + N_\psi/2$.  In the special case $N_\psi=0$, this counting rule reduces to counting powers of momentum $p$ in the \emph{numerator}.
\end{itemize}
These rules are not linearly independent; the 6 power counting rules excluding $R_{4\pi}$ are related by constraint Eq.~(\ref{11c}) in $\d$ dimensions.  The independent rules, which apply {\it simultaneously}, are
\begin{itemize}
\item[--] The coupling constant rules $R_{g,y,\lambda}$.
\item[--] The $\Lambda$ rule $R_\Lambda$.
\item[--] The $4\pi$ counting rule $R_{4\pi}^\prime$ (or equivalently $R_{4\pi}$).
\item[--] Either the field counting rule $R_F$ \emph{or} the rule $R_\chi$,
\end{itemize}
and these four rules provide four different pieces of information. None of the counting rules depends on the number of internal lines in the graph, and only the last rule depends on the number of loops in the EFT.

We have concentrated on applying power counting rules in generic EFTs. There can be symmetry considerations that alter the counting rules. For example, in a theory with small $CP$ violation, $CP$ conserving operators obey the usual counting, whereas $CP$ violating operators are suppressed by a small parameter $\epsilon \ll 1$ that governs the size of $CP$ violation. In QCD, flavor symmetry is broken by the quark mass matrix, which transforms as the adjoint of flavor $SU(3)$. Flavor non-singlet operators are suppressed by powers of $m_q$ given by the minimum number of adjoints needed to construct the operator. Operators with non-zero triality cannot be made from tensor products of adjoints, and have zero coefficient. For fermions, chiral symmetry implies that operators which violate chirality by $\Delta \chi$ have a coefficient with $\Delta \chi/2$ factors of a fermion mass or Yukawa coupling~\cite{Manohar:1983md,Jenkins:2013sda}.  In composite Higgs models~\cite{Kaplan:1983fs,Kaplan:1983sm,Banks:1984gj}, there is a small vacuum misalignment parameter $\epsilon \sim v/f$, where $v \sim 246$\,GeV is the electroweak scale, and $f$ is the analog of $f_\pi$ for the strong dynamics which generates the composite Higgs. This small parameter can be included in the counting of higher dimension operators in these models~\cite{Alonso:2012jc,Alonso:2012px,Alonso:2012pz,Gavela:2014vra,Alonso:2014wta,Hierro:2015nna}, analogous to the way small symmetry breaking parameters are included.

In EFTs, field redefinitions can be used to redefine operators, which is related to using the equations of motion (EOM)~\cite{Politzer:1980me} in $S$-matrix elements. Since the Lagrangian obeys NDA, its variation, which gives the EOM, is also compatible with NDA. For example, the schematic form of the SM Higgs doublet $H$ EOM is 
\begin{align}
D^2 H + m^2 H + \lambda (H^\dagger H) H  + y \overline \psi \psi\,
&=0 ,
\end{align}
and all three terms scale homogeneously as $\Lambda/(4\pi)$ in NDA.  However, the terms are not all of the same order for the other counting rules $R_i$. The values of $(N_F,N_\chi,N_\lambda,N_y)$ for the four terms are $(1,2,0,0)$, $(1,0,0,0)$, $(3,0,1,0)$ and $(2,1,0,1)$, respectively. One could restore the field and $N_\chi$ counting rules by assigning $N_F=-1,N_\chi=1$ to $y$, $N_F=-2, N_\chi=2$  to $\lambda$, and $N_\chi=1$ to $m$, but we will not follow this path, since one loses information by coalescing independent expansions.

The counting of Ref.~\cite{Buchalla:2013eza}, in which fermion fields have chiral dimension $p^{1/2}$, and $y$ has dimension $p$ is equivalent to using the counting parameter
\begin{equation}
\begin{aligned}
\widetilde N_\chi &\equiv N_p + \frac{N_\psi}{2} + N_g + N_y + 2 N_\lambda \\
&\equiv N_\chi + N_g + N_y + 2 N_\lambda\,.
\label{36}
\end{aligned}
\end{equation}
This combination was called the chiral dimension in Ref.~\cite{Buchalla:2014eca} and is compatible with the EOM. Eq.~(\ref{36}) for the chiral 
dimension is closely related to earlier work in chiral perturbation theory, where the charge $Q$ was counted as $\mathcal{O}(p)$ in the chiral counting. Some drawbacks of this counting are discussed in Sec.~\ref{sec:chpt}.

\section{Examples} \label{sec:eg}

In this section, we consider a number of examples which illustrate the EFT power counting rules.
\subsection{Wavefunction Graph}

\begin{figure}[h!]
\includegraphics[width=3.5cm]{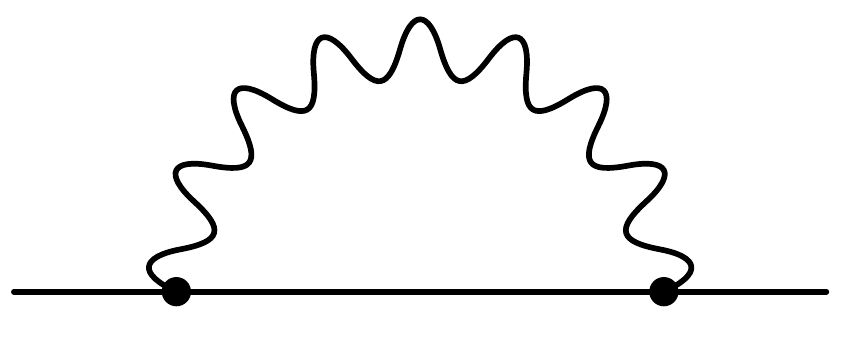}
\caption{\label{fig:1} \it Wavefunction Graph}
\end{figure}
As an example of the counting rules, consider the fermion wavefunction graph in Fig.~\ref{fig:1} in $\d=4$. The two interaction vertices are $g \overline \psi \slashed{A} \psi$, each with 
$N_g=1$, $N_y =0$, $N_\lambda=0$, $N_\phi=0$, $N_A=1$, $N_\psi=2$,  $N_p=0$, $N_\Lambda=0$, $N_{4\pi}=0$, and the loop graph gives $\sim (g^2/(16\pi^2)) \overline \psi \, i\slashed{\partial} \,\psi$ which has 
$N_g=2$, $N_y =0$, $N_\lambda=0$, $N_\phi=0$, $N_A=0$, $N_\psi=2$,  $N_p=1$, $N_\Lambda=0$, $N_{4\pi}=-2$. One can verify that all the counting rules $R_i$ are satisfied.  The NDA master formula Eq.~(\ref{fnda}) gives
\begin{align}
&f^{2} \Lambda^{2} \left[\frac{p}{\Lambda}\right] \left[\frac{\psi}{ f \sqrt{\Lambda}} \right]^{2}  \left[\frac{f g}{ \Lambda }  \right]^{2}
\sim  \frac{g^2}{(4\pi)^2} \psi^2\, p
\label{nda24-1}
\end{align}
which agrees with the final amplitude.

\subsection{Chiral Perturbation Theory}\label{sec:chpt}
The low-energy dynamics of the pion octet in QCD is given by chiral perturbation theory. Conventionally, the leading order  (or order $p^2$) Lagrangian including electromagnetism is
\begin{align}
\LL_2 &= \frac{f^2}{4} \tr \left[ D_\mu U^\dagger D^\mu U + U^\dagger \chi + \chi^\dagger U \right]
\label{l2}
\end{align}
where $U(x)=\exp 2 i \vpiEW(x)/f$, $\chi$ is proportional to the quark mass matrix,  $f$ is the pion decay constant, and $D_\mu U= \partial_\mu + i e A_\mu \left[U,Q\right]$ is the electromagnetic covariant derivative, and $Q=\text{diag}(2/3,-1/3,-1/3)$ is the quark charge matrix. The
NLO (or order $p^4$) Lagrangian is
\begin{align}
\LL_4 &= L_1 \left[ \tr D_\mu U^\dagger D^\mu U \right]^2  \nn
&+ L_4 \left[\tr D_\mu U^\dagger D^\mu U \right]
\left[ \tr U^\dagger \chi + \chi^\dagger U \right]  \nn
& + L_6 \left[ \tr U^\dagger \chi + \chi^\dagger U \right]^2  \nn
& -i L_9 e F^{\mu \nu} \left[ \tr  Q D_\mu U D_\nu U^\dagger  + \tr Q D_\mu U^\dagger D_\nu U \right] \nn
&+ L_{10} e^2  F^{\mu \nu} F_{\mu \nu} \tr U^\dagger  Q  U Q \nn
&+ 2 H_1 e^2  F^{\mu \nu} F_{\mu \nu} \tr  Q^2 + \ldots
\label{l4}
\end{align}
where $F_{\mu\nu}$ is the electromagnetic field-strength tensor,  $L_i$ and $H_i$ are low-energy constants, and we have shown typical terms. The complete $\LL_4$ Lagrangian is given in Ref.~\cite{Pich:1995bw}, and we have simplified the terms in Eq.~(\ref{l4}) to the case of only QED gauge fields.

First consider the chiral limit $\chi \to 0$ with the gauge interactions turned off. The pion field enters in the combination $U(x)=\exp 2 i \vpiEW(x)/f$ by chiral invariance. $U$ obeys the power counting rule of Eqs.~(\ref{nda4},\ref{fnda4})
treating it  as a dimensionless object of order unity, since every power of $\vpiEW$ comes with a factor of $1/f$.  
Since $U(x)$ is dimensionless, the chiral Lagrangian is an expansion in powers of $\partial/\Lambda$, and $N_\chi$ and $N_\Lambda$ counting are identical.

The pion kinetic term in $\LL_2$ has the size given by NDA. NDA implies that the interaction terms $L_{1-8}$ in $\LL_4$  have size $L_i \sim \widetilde L_i/(16\pi^2)$, where $\widetilde L_i$ are order unity~\cite{Manohar:1983md,Manohar:1996cq}. $L_i$ are then of order $10^{-2} - 10^{-3}$, which is the case experimentally (see Table 1 in  Ref.~\cite{Pich:1995bw}). The advantage of  NDA normalization is that the Lagrangian coefficients are order unity.

The mass matrix $\chi$ of $\chi$PT is of order $\chi \sim \Lambda m_q$, where $m_q = \text{diag}(m_u,m_d,m_s)$ is the quark mass matrix, so that $M^2 \sim \Lambda m_q$, where $M$ is the meson mass. This gives a good estimate of the meson masses using known values of the light quark masses. It is conventional in $\chi$PT to treat $\tr \chi U^\dagger + \text{h.c.}$ as order $p^2$. This is convenient for organizing the computation, and also because on-shell mesons have $p^2=M^2$. $p$ and $m_q$ are, however, independent parameters. In $\pi\pi$ scattering, the two-derivative term gives a scattering amplitude of order $E_{\text{CM}}^2/f^2$, where $E_{\text{CM}}$ is the center-of-mass energy, whereas the mass term gives an amplitude of order $M^2/f^2$. The relative importance of the two depends on the kinematics of interest.
One can use $SU(3)$ $\chi$PT, where the kaon is treated as a light particle, or $SU(2)$ $\chi$PT, where the kaon is treated as heavy and integrated out.

The electromagnetic terms are of order $L_{9} \sim e/(16 \pi^2)$ and $L_{10},H_1 \sim e^2/(16\pi^2)$, using
Eqs.~(\ref{nda4},\ref{fnda4}). The $\pi^+-\pi^0$ mass difference, arising from
\begin{align}
\LL &=  c \left( \frac{e}{4\pi} \right)^2 f^2 \Lambda^2 \tr Q U Q U^\dagger
\label{massdiff}
\end{align}
where $c \sim 1$ with NDA normalization, is
\begin{align}
M_{\pi^+}^2-M_{\pi^0}^2 \sim \frac{\alpha}{4\pi} \Lambda^2
\label{massdiffa}
\end{align}
which is the size of the experimentally measured mass difference.

Again, it is conventional to include $L_{9,10},H_1$ as part of the $p^4$ Lagrangian since they are generated along with $L_{1-8}$ by one-loop graphs with $\LL_2$ vertices. However, it is important to remember that this is for convenience in organizing the calculation, and that these terms are not literally of order $p^4$.
A commonly used procedure in $\chi$PT with electromagnetism is to use $p$ power counting with~\cite{Urech:1994hd}
\begin{align}
Q & \sim \mathcal{O}(p), & A_\mu \sim \mathcal{O}(1)\,,
\end{align}
where $Q$ is the charge. This is a useful rule in the context in which it was proposed. However, it should not be taken as a fundamental principle to be blindly applied in all cases. For example, the long distance (i.e.\ $p \to 0$) Coulomb field of a particle is proportional to the electric charge $Q$, and so $Q$ certainly does not vanish at zero momentum. 

The power counting rules discussed here agree with the conventional ones used in $\chi$PT, with the distinction that we are treating Lagrangian parameters such as $m_q$ and $\alpha$ as independent parameters, since they are not equal to kinematic variables such as $p$. This distinction is particularly important in the case of HEFT discussed in Sec.~\ref{sec:heft}, where one is investigating the dynamics of a high energy theory which is unknown.

\subsection{Low Energy Weak Interactions}\label{sec:fermi}

The Fermi theory of weak interactions is the low-energy limit of the electroweak sector of the SM theory. It provides a well-known pedagogical example of an EFT arising from a weakly coupled theory, and shows how low-energy experiments can be used to determine the pattern of coefficients of EFT operators and to deduce the UV theory. Historically, the process took several decades.

The structure of the weak interactions is surprisingly subtle.
At tree-level, single $W$ exchange gives the interaction term
\begin{align}
\LL &= -\frac{4 G_F}{\sqrt 2}V_{ij} V_{kl}^* \left( \overline \psi_i \gamma_\mu P_L \psi_j \right) \left( \overline \psi_l \gamma^\mu P_L \psi_k \right)
\label{1.1}
\end{align}
in the quark sector, where $i,j,k,l$ are flavor indices, and a similar term in the lepton sector. The SM gauge symmetry is spontaneously broken by the Higgs mechanism, leading to massive charged gauge bosons with mass $M_W =g_2 v/2$, and a massive scalar with mass $M_H = \sqrt{2 \lambda}\, v$. The Fermi constant is
\begin{align}
\frac{4G_F}{\sqrt{2}} &= \frac{g_2^2}{2M_W^2}= \frac{2}{v^2}
\label{1.3}
\end{align}
In the weak coupling limit $g_2 \to 0$, $M_W \to 0$, but $G_F$ remains fixed. As far as the EFT is concerned, the only relevant quantity is $G_FV_{ij} V_{kl}^*$; $g_2$ and $M_W$ are parameters of the UV theory.

There are also $\Delta S=2$ interactions,
\begin{align}
\LL^{(\Delta S=2)} &= -\frac{4 G_F}{\sqrt 2} C_2  \left( \overline d \gamma_\mu P_L s\right) \left( \overline d \gamma^\mu P_L s \right)
\label{1.2}
\end{align}
which are generated at one-loop in the Standard Model.  The coefficient $C_2$, generated by box graphs, depends at one-loop on the CKM angles $V_{is}V_{id}^*$, as well as $M_W$ and quark masses. Higher order corrections also depend on $M_H$.

It is clear that the low-energy weak interactions are not, in any sense, described by a ``one-scale, one-coupling'' theory~\cite{Panico:2015jxa}. The CKM elements $V_{ij} V_{kl}^*$ vary over five orders of magnitude, and the flavor structure of the weak interactions is a crucial part of the Standard Model. Any model of new physics has to incorporate the  SM flavor structure if it is to be compatible with experiment. Even neglecting flavor, the SM has at least two dimensionless couplings $g_2$ and $\lambda$, and radiative corrections depend on these, either directly through vertex couplings, or indirectly, through the $M_W/M_H$ mass ratio.

Let us now pretend that we do not know the underlying $SU(2) \times U(1)$ electroweak theory, and see what we can learn from EFT methods. This is not an esoteric exercise --- it was the way in which the structure of the weak interactions was determined historically.  We will concentrate on the lepton sector, to avoid the additional subtleties of flavor and the GIM mechanism~\cite{Glashow:1970gm}, and also to not worry about differences between $M_W$ and $M_Z$, to simplify the discussion. 

The leptonic version of Eq.~(\ref{1.1}) determines $G_F$ via the well-known formula for the muon decay rate,
$\Gamma= G_F^2 m_\mu^5/(192 \pi^3)$. This simple form follows because of two accidents, that neutrino mass differences are small, and determining neutrino flavors is difficult experimentally. Otherwise, the muon decay rate would be broken up into individual neutrino flavor states, with factors of the PMNS matrix which can vary by several orders of magnitude.

From Eq.~(\ref{nda4}), we get the inequality
\begin{align}
G_F \alt \frac{(4\pi)^2}{\Lambda^2} \implies \Lambda \alt 4 \pi v
\label{1.4}
\end{align}
since we have made no assumption about whether the weak interactions are weakly or strongly coupled. It is worth repeating that a measurement of $G_F$ does not determine $\Lambda$, but only gives an inequality Eq.~(\ref{1.4}).
Assuming that Eq.~(\ref{1.1}) is generated by tree-level gauge boson exchange in the UV theory allows us to obtain $g_2/M_W$, but not $g_2$ or $M_W$ separately. For example, the Abbott-Farhi model~\cite{Abbott:1981re,Abbott:1981yg}, a strongly interacting model for the weak interactions, has a different $W$ mass than the SM.  

Additional experimental information is needed to obtain $\Lambda$, which sets the scale of the momentum expansion. 
One method is to use neutral current neutrino scattering experiments to determine $\sin^2\theta_W$, which, when combined with $\alpha$ gives $g_2$. Then we can separately obtain $g_2$ and $M_W$ using Eq.~(\ref{1.3}). A more direct method is to use energy dependence of parity violating electron scattering through $\gamma-Z$ interference, which shows that the scale $\Lambda$ of the momentum expansion is $M_Z \sim M_W$. In the EFT, this is due to the
$p^2/M_Z^2$ operator obtained by expanding $1/(p^2-M_Z^2)$ in powers of $p$.  In other words, $\Lambda \sim M_Z \sim M_W$ can be determined from low-energy measurements by comparing the coefficient of the dimension-six $\psi^4$ and dimension-eight $\partial^2 \psi^4$ operators in the EFT.

With this additional piece of information, we see that Eq.~(\ref{1.4}) is an inequality, so that the SM is the low-energy limit of a weakly coupled theory. Assuming that the muon decay operator is suppressed by one power of the UV coupling $g_2^2$, Eq.~(\ref{nda4}) implies that the operator is\footnote{If we assume that EFT operators have power of $g,y,\lambda$ of the UV theory through matching, then we count those powers in the same way as the EFT couplings, Eq.~(\ref{nda}).}
\begin{align}
\LL \sim \frac{(4\pi)^2}{\Lambda^2} \left( \frac{g_2}{4\pi} \right)^2 \psi^4
\label{1.5}
\end{align}
so that $G_F \sim g_2^2/M_W^2$. This relation then allows us to estimate $g_2$ and $M_W$ separately. In the SM, not all four-fermion operators arise at the same order in $g_2$. The $\Delta S=2$ operator is not of the size Eq.~(\ref{1.5}), but has an additional suppression of two powers of the coupling,
\begin{align}
\LL^{(\Delta S=2)} \sim \frac{(4\pi)^2}{\Lambda^2} \left( \frac{g_2}{4\pi} \right)^4 \psi^4 \sim G_F \frac{\alpha_W}{4\pi} \psi^4
\label{1.6}
\end{align}
If we had assumed (incorrectly) that the $\Delta S=2$ Lagrangian had the form Eq.~(\ref{1.5}), we would have obtained the wrong value for $g_2$.

The pattern of suppression is \emph{not} a prediction of the EFT, but must be determined experimentally. It depends on the underlying UV theory, but can be determined experimentally solely by using low-energy measurements that can be computed using the EFT, as was done for the SM. 
The process can be quite involved --- it took several decades to understand
the electroweak sector of the SM, and much of this work was done using low-energy measurements, before the advent of colliders energetic enough to produce the $W$ and $Z$.

\subsection{Running of the Low Energy Weak Interactions}\label{sec:running}
The renormalization group equations for the low-energy weak interation Lagrangian also provides a nice example of the utility of the $4\pi$ counting rules of NDA. We will concentrate on just two operators, a four-fermion operator $O_1 \sim \psi^4$, and a magnetic moment operator $O_7 \sim m \psi^2 X$, and show how the EFT power counting rules are already implicit in the standard form of the weak Lagrangian for $b \to s$ transitions~\cite{Grinstein:1990tj},
\begin{align}
\begin{aligned}
\LL_W = &-\frac{4 G_F}{\sqrt 2}V_{cb} V_{cs}^* \sum_i C_i\, O_i\,,\\
O_1 &= \left(\overline c \gamma^\mu P_L b\right)\left(\overline  s \gamma_\mu P_L c\right)\,,\\
O_7 &= \frac{e}{16\pi^2}  \left(\overline s \sigma^{\mu \nu}  P_R b\right)\, F_{\mu \nu} \, m_b\,.
\end{aligned}
\label{36.1}
\end{align}
The NDA form of the Lagrangian from Eq.~(\ref{nda4}) gives precisely this relative factor of $e/(16\pi^2)$  between $O_7$ and $O_1$.
The renormalization group evolution equations have a homogenous form when the operators are normalized as in Eq.~(\ref{36.1}) as  expected, since NDA is consistent with renormalization. In Ref.~\cite{Grinstein:1990tj}, the $e/(16 \pi^2)$ factor was introduced to make the evolution equations have homogeneous form. There is an analogous NDA scaling that was used in Ref.~\cite{Jenkins:2013sda} to simplify the form of the evolution equations for the SMEFT, which has eight different operator classes of dimension-six with their individual scaling factors.

\subsection{SMEFT}\label{subsec:smeft}

SMEFT is another instructive example of the use of Eq.~(\ref{nda}). The SM and SMEFT lepton and baryon number preserving operators of mass dimension $d \le 6$ and chiral number $N_\chi = N_p + N_\psi/2$ are shown schematically in Table~\ref{tab:ops} using NDA normalization. $N_\chi$ reflects the different $4\pi$ weights of operators with a given mass dimension $d$.\footnote{ It is related to the NDA weight $w$ defined in Ref.~\cite{Jenkins:2013sda}. As shown in  Ref.~\cite{Jenkins:2013sda}, $w$ explains the pattern of the one-loop SMEFT anomalous dimensions~\cite{Grojean:2013kd,Elias-Miro:2013gya,Jenkins:2013zja,Jenkins:2013wua,Alonso:2013hga,Alonso:2014zka}, as well as the approximate holomorphy of the one-loop anomalous dimension matrix for $d=6$ operators found in Ref.~\cite{Alonso:2014rga}, and proven in a more general context in Ref.~\cite{Elias-Miro:2014eia,Cheung:2015aba}.}

\begin{table}
\renewcommand{\arraystretch}{1.5}
\begin{align*}
\begin{array}{c|c|c|c}
 \text{Operator} & \text{$d$} & N_\chi  &  \text{NDA Form} \\
 \hline
  \hline
H^2 & 2  & 0  & \Lambda^2 H^2  \\
\psi^2 & 3 & 1  & \Lambda \psi^2 \\
H^4  & 4 & 0  & (4\pi)^2\, H^4  \\
\psi^2 H & 4 & 1 & (4 \pi) \, \psi^2 H  \\
\psi^2 D  & 4 &  2 &\psi^2 D  \\
H^2 D^2  &  4 & 2 & H^2 D^2  \\
X^2 & 4 & 2 & X^2  \\
\hline
H^6 & 6 & 0  & \frac{(4\pi)^4}{\Lambda^2}\, H^6  \\
\psi^2 H^3 & 6 & 1  &\frac{(4\pi)^3}{\Lambda^2}\, \psi^2 H^3  2  \\
H^4 D^2 & 6 & 2  & \frac{(4\pi)^2}{\Lambda^2}\, H^4 D^2 \\
X^2 H^2 & 6 & 2 & \frac{(4\pi)^2}{\Lambda^2}\, X^2 H^2   \\
\psi^2 X H & 6 & 2  & \frac{(4\pi)^2}{\Lambda^2}\, \psi^2 X H    \\
\psi^2 H^2 D & 6 &  2 & \frac{(4\pi)^2}{\Lambda^2}\, \psi^2 H^2 D   \\
\psi^4 & 6 & 2 &  \frac{(4\pi)^2}{\Lambda^2}\, \psi^4\\
X^3 & 6 & 3  &  \frac{(4\pi)}{\Lambda^2}\, X^3 
\end{array}
\end{align*}
\caption{{\it SM and SMEFT lepton and baryon number preserving operators of mass dimension $d \le 6$ and chiral number $N_\chi \equiv N_p + N_\psi/2$, normalized using NDA rule Eq.~(\ref{nda4}) in $\d=4$ spacetime dimensions.  The notation is schematic, with $H$ the Higgs field, $\psi$ a fermion field, $X_{\mu \nu}$ a field-strength tensor and $D$ a covariant derivative. All indices are suppressed. The complete set of operators can be found in Refs.~\cite{Buchmuller:1985jz,Grzadkowski:2010es}.} \label{tab:ops}}
\end{table}

The SMEFT Lagrangian is obtained from the SM Lagrangian by adding all operators of mass dimension $d>4$ constructed from SM fields with suppressions of $1/\Lambda^{d-4}$ in $\d=4$ spacetime dimensions.  Schematically, the SMEFT Lagrangian (ignoring lepton and baryon number violating operators) is
\begin{equation}
\begin{split}
\LL \sim& -\frac14 X^2 + \overline \psi i\slashed{D} \psi + D_\mu H^\dagger D^\mu H \\
& - m_S^2 H^\dagger H  - y \overline \psi \psi H - \lambda (H^\dagger H)^2 \\
& + C_H \frac{H^6}{\Lambda^2}   + C_{\psi^2 H^3} \frac{\psi^2 H^3}{\Lambda^2} + C_{\psi^4} \frac{\psi^4}{\Lambda^2} + \ldots
\end{split}
\label{41}
\end{equation}
where we have used the conventional normalization of coefficients, and we have explicitly shown only a few dimension-six operators. The same Lagrangian in NDA normalization is given using Eq.~(\ref{nda4}),
\begin{align}
\LL \sim& -\frac14 X^2 + \overline \psi i\slashed{D} \psi + D_\mu H^\dagger D^\mu H \nn
&  - \widehat m_S^2 \left[\Lambda^2 H^\dagger H \right] - \widehat y \left[  4\pi\, \overline \psi \psi H \right]  - \widehat \lambda \left[ (4\pi)^2 (H^\dagger H)^2 \right] \nn
& + \widehat C_H \frac{(4\pi)^4 H^6}{\Lambda^2}   + \widehat C_{\psi^2 H^3} \frac{(4\pi)^3 \psi^2 H^3}{\Lambda^2} + \widehat C_{\psi^4} \frac{(4\pi)^2 \psi^4}{\Lambda^4} + \ldots
\label{42}
\end{align}
The two Lagrangians are \emph{identical}, so the couplings are related by
\begin{equation}
\begin{aligned}
\widehat m_S &= \frac{m_S}{\Lambda}, & \widehat y &= \frac{y}{4\pi}, & \widehat \lambda &= \frac{\lambda}{(4\pi)^2}, \\
\widehat C_H &= \frac{C_H}{(4\pi)^4}, & \widehat C_{\psi^2 H^3} &= \frac{C_{\psi^2 H^3}}{(4\pi)^3}, &
\widehat C_{\psi^4} &= \frac{C_{\psi^4}}{(4\pi)^2}.
\end{aligned}
\label{43}
\end{equation}
\begin{figure*}
\renewcommand{\arraycolsep}{0.5cm}
\begin{align*}
\begin{array}{ccc}
\lower0.2cm\hbox{\includegraphics[scale=0.25]{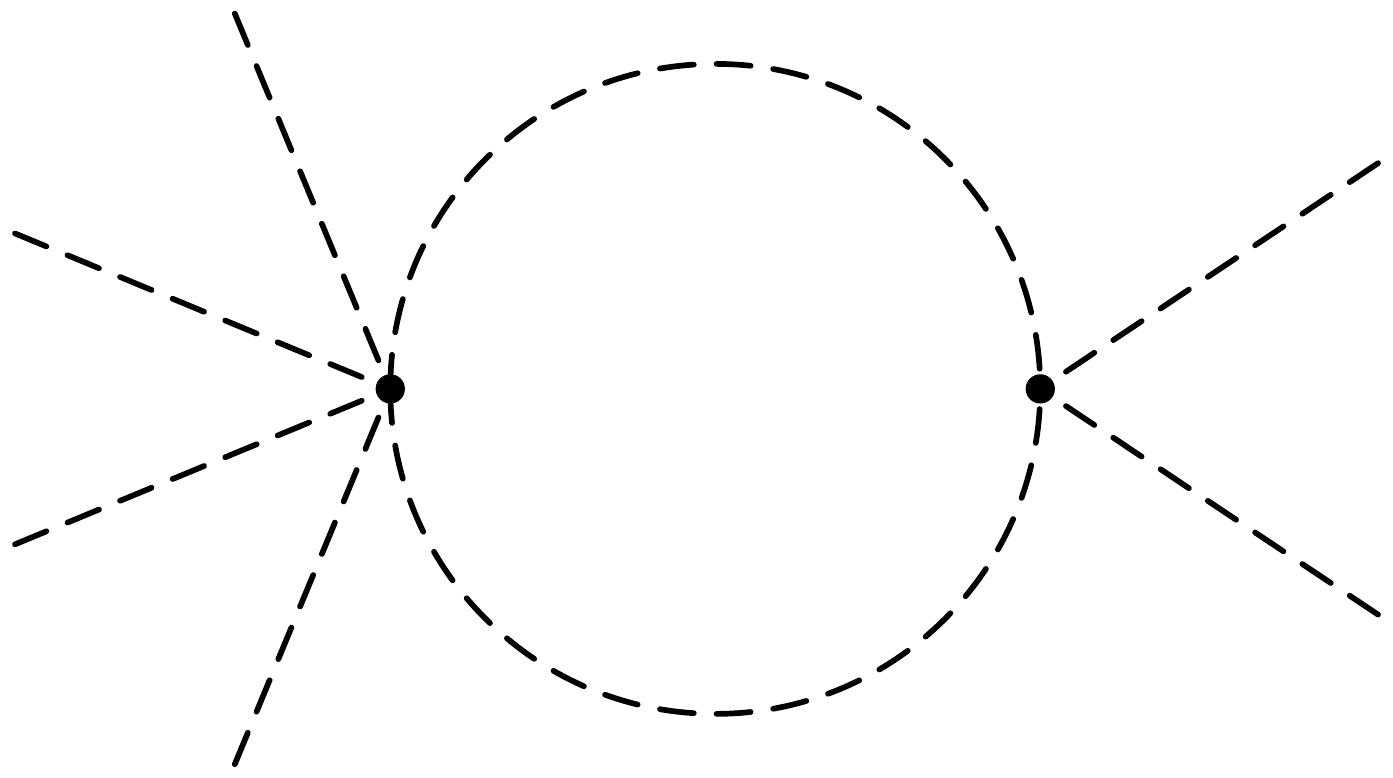}} & \includegraphics[scale=0.25]{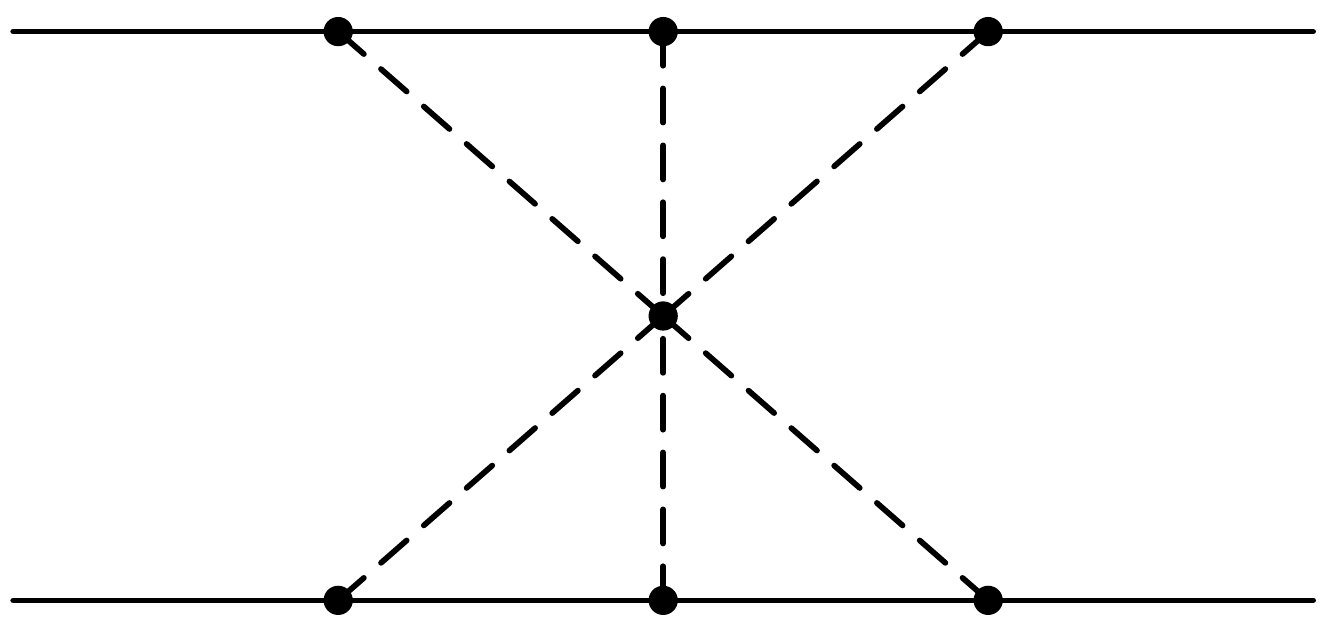} & \includegraphics[scale=0.25]{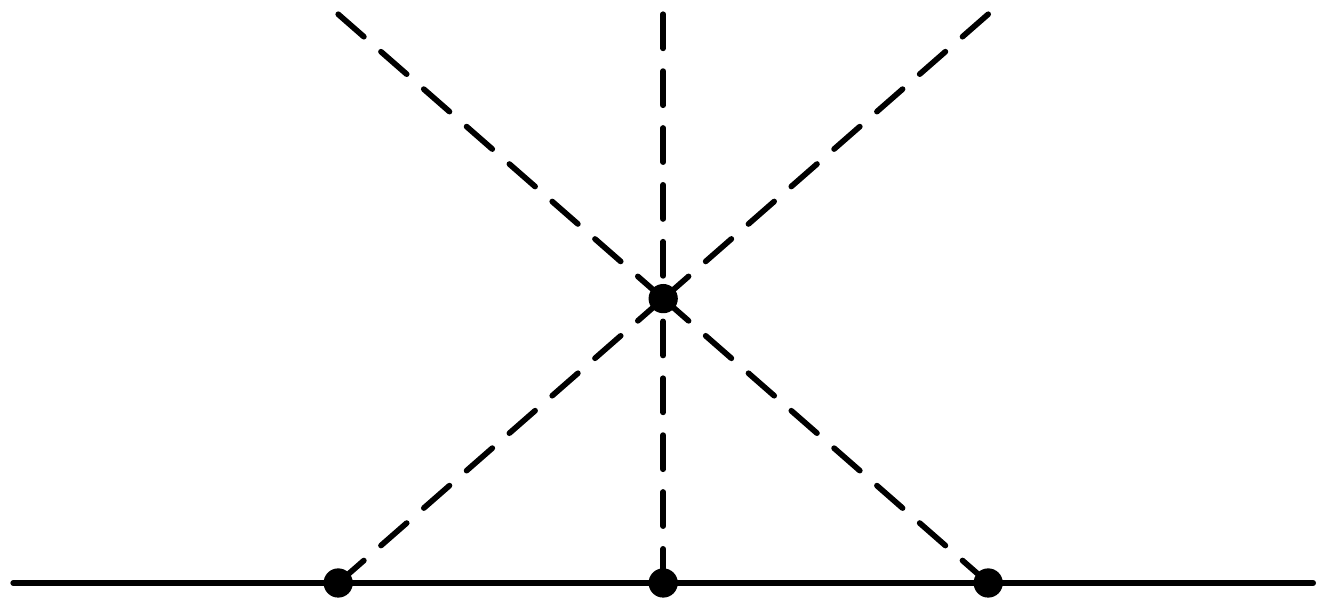} \\
(a) & (b) & (c)  \\
\end{array}
\end{align*}
\begin{align*}
\begin{array}{cc}
\raise0.2cm\hbox{\includegraphics[scale=0.25]{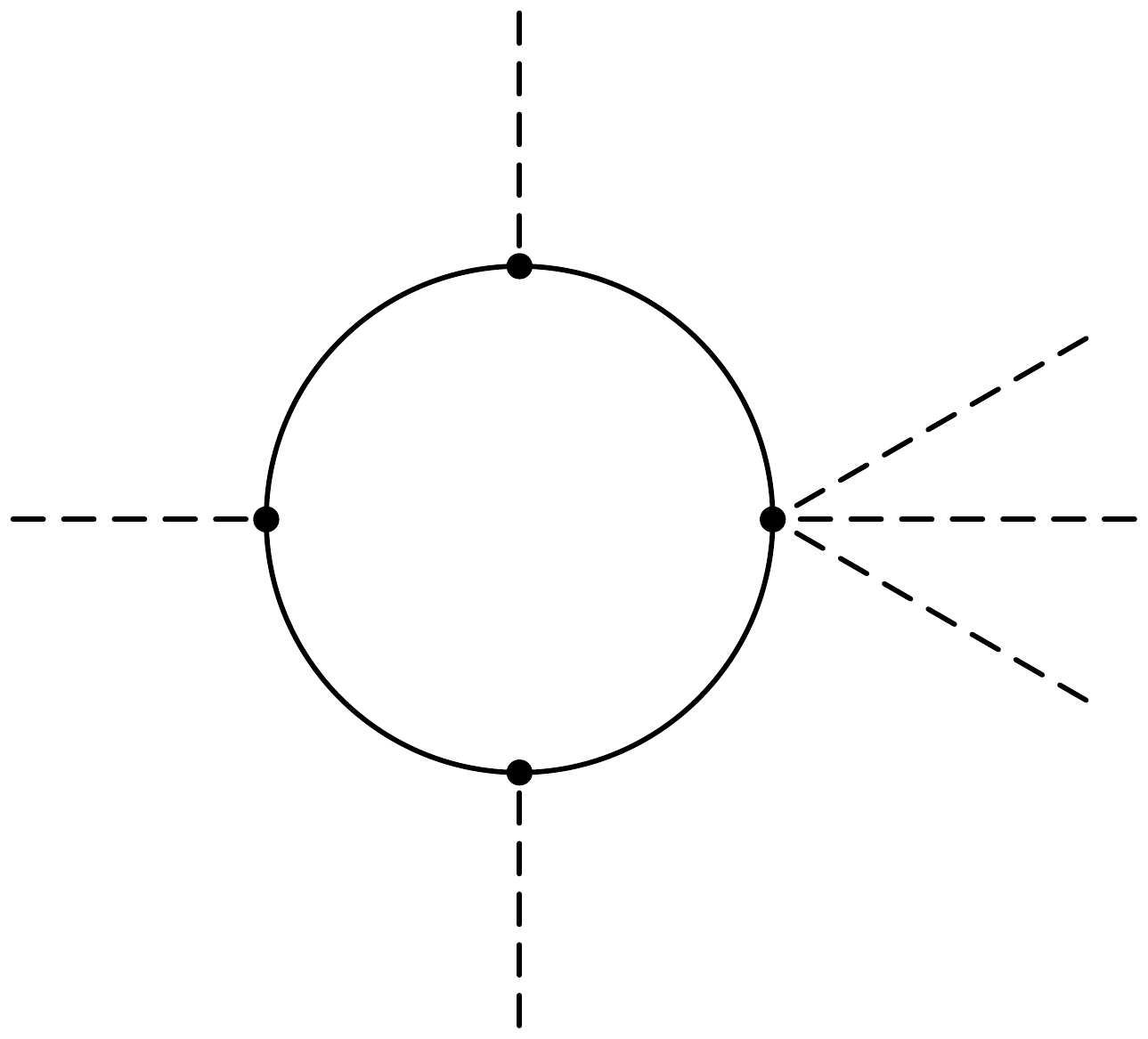}} &  \includegraphics[scale=0.25]{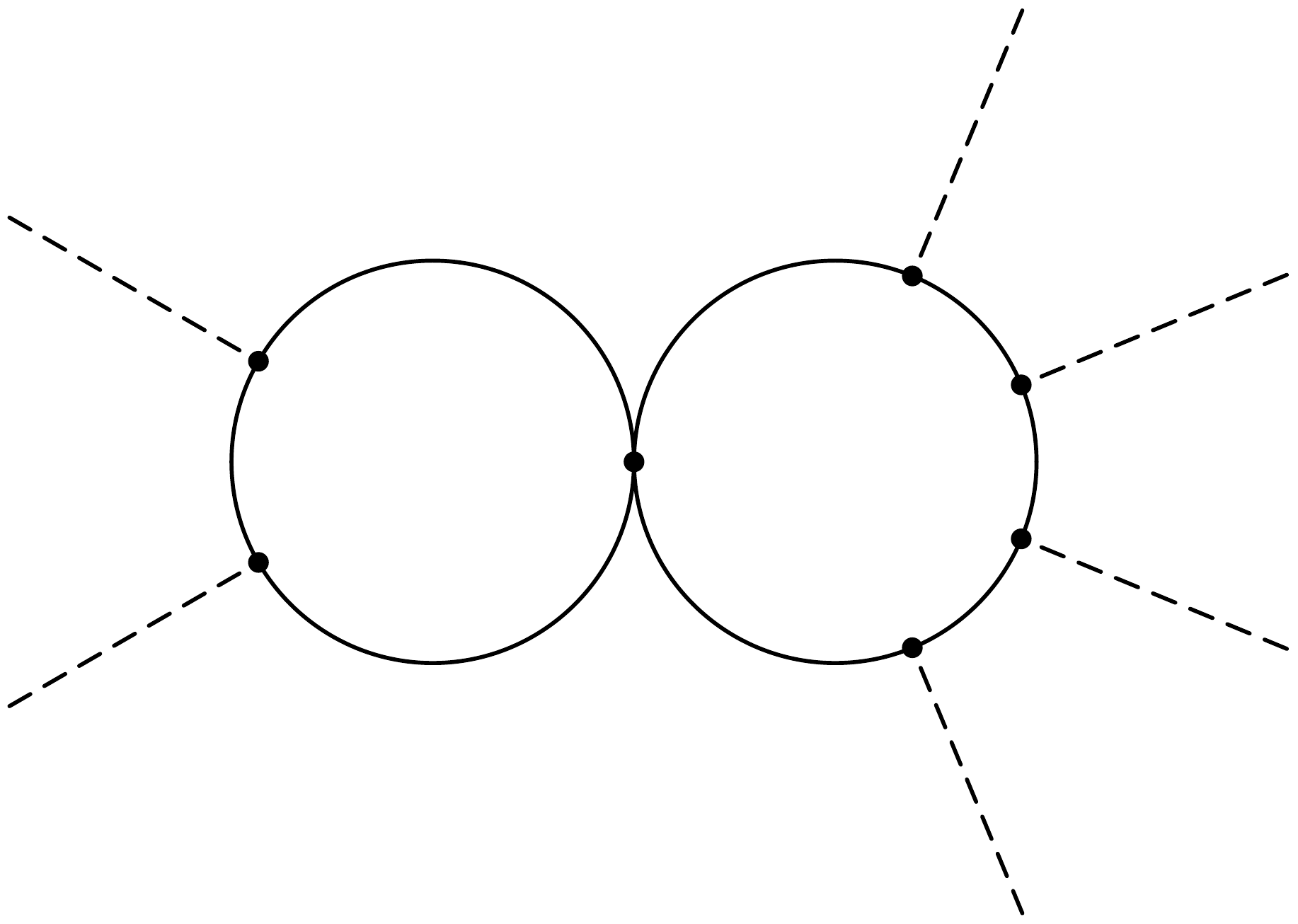} \\
(d) & (e) \end{array}
\end{align*}
\caption{\label{fig:2} \it A few SMEFT graphs. $H$ is denoted by a dashed line, and $\psi$ by a solid line.}
\end{figure*}

Now consider a few sample amplitudes shown in Fig.~\ref{fig:2}. Graphs $(a)-(e)$ give the contributions
\begin{equation}
\begin{aligned}
\delta C_H & \sim \frac{1}{16\pi^2} \lambda\, C_H ,\\
\delta C_{\psi^4} & \sim \frac{1}{(16\pi^2)^4} y^6\, C_H ,\\
\delta C_{\psi^2 H^3} & \sim \frac{1}{(16\pi^2)^2} y^3\,  C_H ,\\
\delta C_H & \sim \frac{1}{16\pi^2} y^3\, C_{\psi^2 H^3} ,\\
\delta C_H & \sim \frac{1}{(16\pi^2)^2} y^6\,  C_{\psi^4},
\end{aligned}
\label{44}
\end{equation}
respectively, using the Lagrangian Eq.~(\ref{41}), and using $1/(16\pi^2)$ as an estimate for each loop. If instead we use the NDA form Eq.~(\ref{42}) and the rescaling Eq.~(\ref{43}), it follows that
\begin{equation}
\begin{aligned}
\delta \widehat C_H & \sim  \widehat  \lambda\, \widehat C_H ,\\
\delta \widehat C_{\psi^4} & \sim  \widehat y^6\, \widehat C_H , \\
\delta \widehat C_{\psi^2 H^3} & \sim \widehat y^3\, \widehat C_H ,\\
\delta \widehat C_H & \sim \widehat y^3\, \widehat C_{\psi^2 H^3} ,\\
\delta \widehat C_H & \sim  \widehat y^6\, \widehat  C_{\psi^4}\,.
\end{aligned}
\label{45}
\end{equation}
All the $4\pi$ factors have disappeared, and one obtains a very simple form for the amplitudes. To identify the dependence in Eq.~(\ref{44}), it is necessary to draw the diagrams and determine the number of loops. Eq.~(\ref{45}), instead, has a universal form which is independent of the graphs,
\begin{align}
\delta\widehat C_i & \sim \prod_{k} \widehat C_{i_k}\,.
\label{46}
\end{align}
Note that no assumption has been made that the theory is strongly coupled. The results are equally valid for strongly and weakly coupled theories. 

The NDA form Eq.~(\ref{45}) also shows that in strongly coupled theories $\widehat C \lesssim 1$~\cite{Manohar:1983md}. The reason is that if $\widehat C \gg 1$, then the hierarchy of equations Eq.~(\ref{46}) is unstable, because higher order contributions to $\widehat C_i$ are much larger than $\widehat C_i$. On the other hand, there is no inconsistency if $\widehat C_i \ll 1$, since all that implies is that higher order corrections are small, a sign of a weakly coupled theory. Eq.~(\ref{46}) also implies that an interaction becomes strongly coupled  when $\widehat C \sim 1$.  For the SM couplings with the conventional normalization, strong coupling is $g \sim 4\pi$, $y \sim 4\pi$ and $\lambda \sim (4\pi)^2$.  For the SMEFT, the dimension six operators are strongly coupled when $C_H \sim (4\pi)^4$, $C_{\psi^2 H^3} \sim (4\pi)^3$, $C_{\psi^2} \sim (4\pi)^2$, etc. Thus, the Lagrangian coefficient with NDA normalization reflects how close the interaction is to its strong coupling value of order unity, with all factors of $4\pi$ absorbed by the normalization.

\subsection{Matching}

Finally, we demonstrate that the NDA form helps in determining the size of EFT coefficients from matching conditions, e.g.\ when integrating out heavy particles.   (See e.g.\ Ref.~\cite{Manohar:1996cq} for a review on how to compute matching in EFTs.)   To illustrate this point, we consider examples of tree and loop matching.

\begin{figure*}
\renewcommand{\arraycolsep}{0.75cm}
\begin{align*}
\begin{array}{ccc}
\includegraphics[scale=0.25]{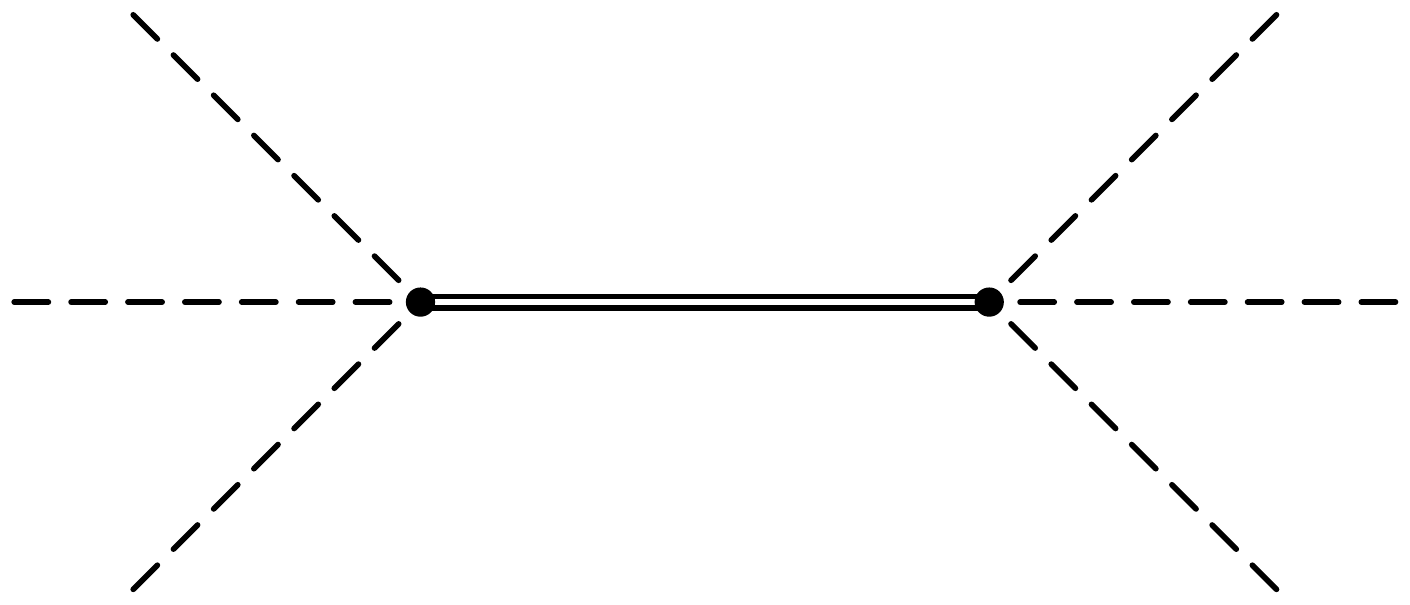} & \includegraphics[scale=0.25]{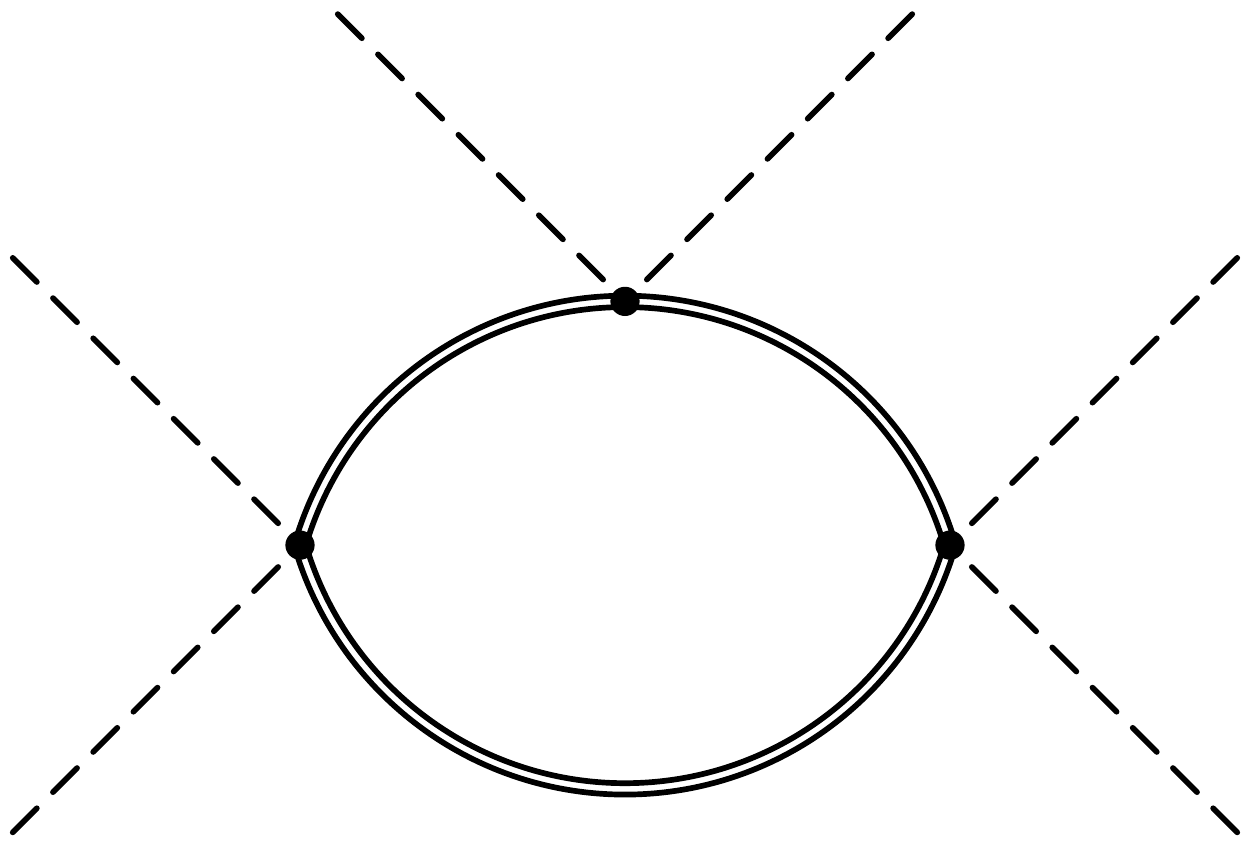} & \lower0.7cm\hbox{\includegraphics[scale=0.25]{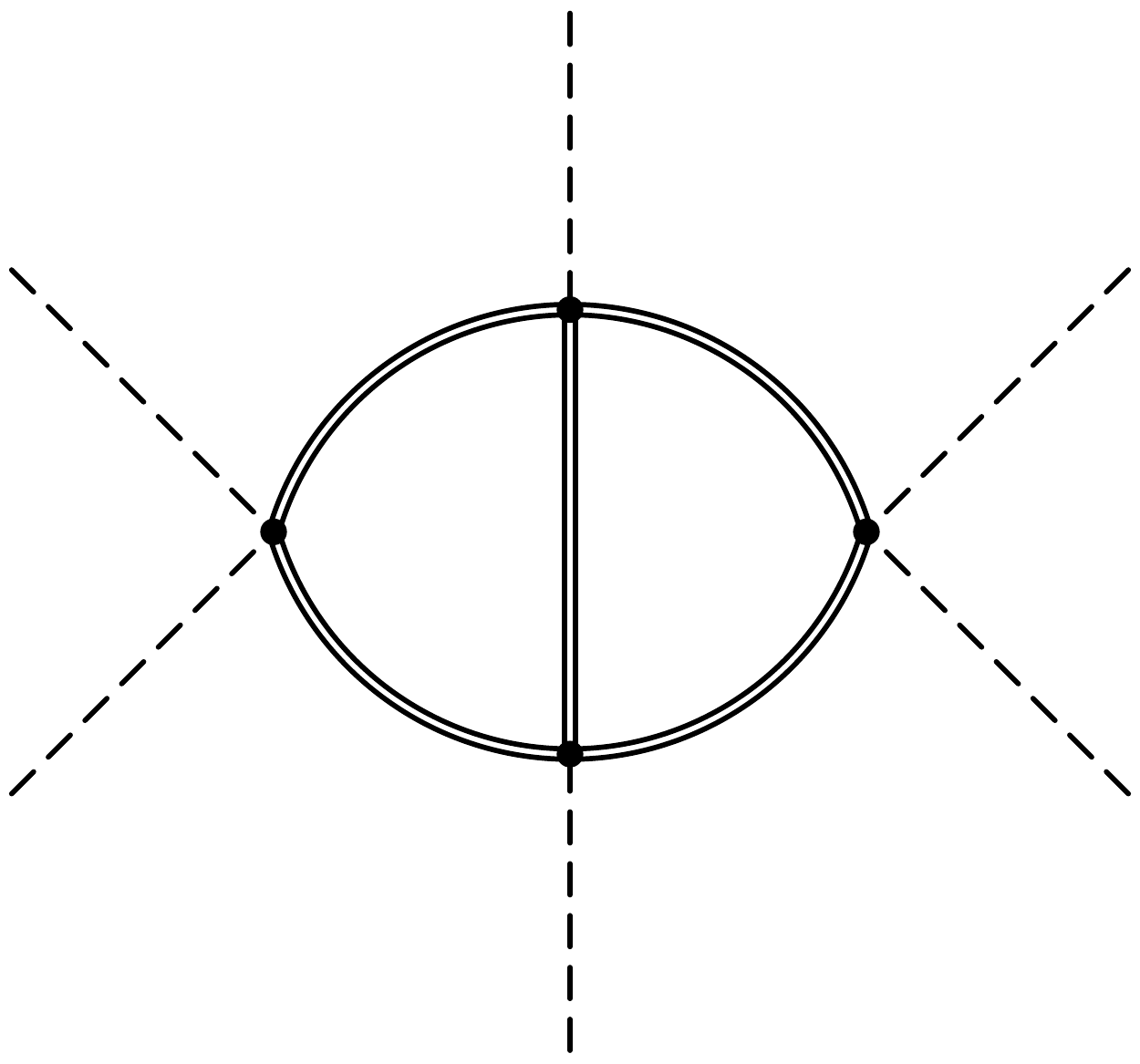}} \\
(a) & (b) & (c)  \\
\end{array}
\end{align*}
\caption{\label{fig:3} \it Graphs contributing to $\phi^6$ matching from integrating out a heavy scalar $\Phi$. $\phi$ is denoted by a dashed line, and $\Phi$ by a double line.}
\end{figure*}

An example of matching a tree diagram is the Fermi theory, where integrating out the electroweak gauge bosons results in four-fermion interactions. The charged current interaction Lagrangian is
\begin{equation}
\begin{split}
\LL_{\text{int}} &=\bar{\psi}_Li\left(\slashed{\partial}+i\,g\,\slashed{W}\right)\psi_L\\
&=\bar{\psi}_Li\left(\slashed{\partial}+i\, 4\pi\,\widehat{g}\,\slashed{W}\right)\psi_L\,,
\end{split}
\end{equation}
where the first and second lines are the conventional and NDA normalizations, respectively, and the couplings are related by $\widehat g = g/(4\pi)$. The charged current four-fermion interactions have Fermi couplings proportional to 
\begin{align}
C_{\psi^4}\sim \frac{g_2^2}{M_W^2}\qquad\text{and }\qquad \widehat C_{\psi^4}\sim \frac{\widehat{g}^2}{M_W^2}\,
\end{align}
in the two normalizations, which are equivalent since $C_{\psi^4}=(4\pi)^2\widehat C_{\psi^4}$.

For examples of loop matching, consider a theory with a light scalar $\phi$ and a heavy scalar $\Phi$ with mass $M$, and interaction terms
\begin{equation}
\begin{split}
\LL_{\text{int}} &= - \lambda_1\, \phi^3\, \Phi - \lambda_2\, \phi^2\, \Phi^2 - \lambda_3\, \phi \, \Phi^3 \\
&= - (4\pi)^2 \left[ \widehat \lambda_1\, \phi^3\, \Phi + \widehat \lambda_2\, \phi^2\, \Phi^2 + \widehat \lambda_3\, \phi \, \Phi^3 \right],
\end{split}
\label{47}
\end{equation}
where again the first line is the conventional normalization, and the second line is NDA normalization. The couplings are related by $\widehat \lambda_i = \lambda_i/(16\pi^2)$. Three sample graphs which produce a $\phi^6$ operator in the theory below $M$ are shown in Fig.~\ref{fig:3}. Since the $\Phi$ propagator is $1/(p^2-M^2)$, we see that $\Lambda$, the scale of the momentum expansion, is fixed to be $M$. The $\phi^6$ operator coefficients $c_6$ given by the three graphs $(a)-(c)$ are 
\begin{align}
c_6 &\sim \frac{\lambda_1^2}{M^2}\,, &
c_6 &\sim \frac{\lambda_2^3}{16 \pi^2 M^2} \,, &
c_6 &\sim \frac{\lambda_2^2 \lambda_3^2}{(16\pi^2)^2 M^2}\,,
\end{align}
respectively.  Using NDA form,
\begin{align}
c_6 \phi^6 &= \widehat c_6 \left[ (4\pi)^4 \phi^6 \right],
\end{align}
so that $\widehat c_6= c_6/(4\pi)^4$, we find that the three graphs give
\begin{align}
\widehat c_6 &\sim \frac{\widehat \lambda_1^2}{M^2}\,, &
\widehat c_6 &\sim \frac{\widehat \lambda_2^3}{M^2} \,, &
\widehat c_6 &\sim \frac{\widehat \lambda_2^2 \widehat \lambda_3^2}{M^2}\,,
\end{align}
and the loop factors have disappeared. Thus, NDA also gives an efficient way to estimate matching conditions with no loop factors.

The examples in this section also show that what is important in a field theory is whether it is strongly or weakly coupled, not the loop factors. The NDA normalization is convenient because all coefficients are expressed as a fraction of their strong coupling value. More non-trivial examples of matching, which obey NDA, have been discussed in Ref.~\cite{Jenkins:2013fya} in the context of minimal coupling, and in Ref.~\cite{Manohar:2013rga} for an exactly solvable model.
\subsection{Gauge Field Strengths}

The counting rules show that gauge field strengths $X_{\mu\nu}$ are normalized as
\begin{align}
\frac{4 \pi X_{\mu \nu}}{\Lambda^2}
\label{48}
\end{align}
in the power counting formula, since $X_{\mu \nu} \sim \partial A$.  This scaling gives a properly normalized gauge kinetic term, and it
gives Lagrangian terms
\begin{align}
&\frac{4 \pi}{\Lambda} \overline \psi \sigma^{\mu \nu} X_{\mu \nu} \psi , & & \frac{4\pi}{\Lambda^2} f_{ABC} X^A_{\mu \nu}X^B_{\nu \lambda}X^C_{\lambda \mu},
\label{50}
\end{align}
for anomalous magnetic moment and triple-gauge interactions, respectively.

If $X_{\mu \nu}$ is an elementary field that couples to particles in the high-energy theory, and the EFT is given by integrating out heavy particles, then one can see graphically that every gauge field comes along with at least one factor of the gauge coupling $g$. The counting rule for $gX_{\mu\nu}$ is
\begin{align}
\frac{g X_{\mu \nu}}{\Lambda^2}\,,
\label{49}
\end{align}
which leads to the form
\begin{align}
&\frac{g}{\Lambda} \overline \psi \sigma^{\mu \nu}  X_{\mu \nu} \psi, & & \frac{g^3}{16\pi^2 \Lambda^2} f_{ABC} X^A_{\mu \nu}X^B_{\nu \lambda}X^C_{\lambda \mu},
\label{51}
\end{align}
for the anomalous magnetic moment and triple-gauge interactions.

In strong coupling theories where $g \sim 4\pi$, the two normalizations Eqs.~(\ref{50}) and (\ref{51}) are equivalent, but they differ in weakly coupled theories. The difference is \emph{not} due to two alternate power counting rules. Rather, it is a dynamical question about the underlying high-energy theory. Gauge invariance does not imply that every $X_{\mu \nu}$ should come with a gauge coupling constant $g$. In EFTs arising from UV theories where $X_{\mu \nu}$ is a fundamental gauge field, interaction terms come with a $g$; however, there is no reason for this form if the gauge boson itself is composite due to some strong dynamics at high energies.

\boldmath
\section{$\Lambda$ vs $p$ Counting for Cross Sections}\label{sec:compare}
\unboldmath

In this section, we compare the $R_\Lambda$ and $R_\chi$ counting rules for experimentally measured quantities such as cross sections. We will use NDA counting in this section, to show how the $4\pi$ counting rules also work out nicely for cross sections. The examples we consider only contain scalar fields, so the $R_\chi$ counting is equivalent to Weinberg's power counting rule for momentum $p$. To avoid confusion, we wish to stress at the outset that \emph{both} $p$ and $\Lambda$ counting rules are valid. The main point is that phase space depends on $p$ but not on $\Lambda$, so one has to include phase space factors when applying $p$ counting rules to cross sections, or equivalently, apply the counting rules to cut graphs which contain additional loops. $\Lambda$ counting does not depend on the number of loops, but $p$ counting does.
For this reason, manifest power counting of cross sections is controlled by $\Lambda$, not by $p$.

Normalize the scattering amplitude to have the NDA form Eq.~(\ref{nda}), and assume the overall power of $\Lambda$ is $N_\Lambda$. The contribution to the $2 \to n$ cross section in $\d$ dimensions from the product of two amplitudes $\mathcal{A}^{(1)}\mathcal{A}^{(2)*}$, neglecting particle masses, has the form
\begin{align}
\sigma \sim&   \pi (4\pi)^{\d/2} E^{2-\d} \!\! \left( \frac{E}{\Lambda}\right)^{-N^{(1)}_\Lambda-N^{(2)}_\Lambda} \hspace{-0.2cm} \left[\frac{g}{(4 \pi)^{\d/4} E^{(4-\d)/2} }  \right]^{N_g^{(1)} + N_g^{(2)}}\nn
& \left[\frac{y}{(4 \pi)^{\d/4} E^{(4-\d)/2} } \right]^{N_y^{(1)} + N_y^{(2)}}\! \left[\frac{\lambda}{(4 \pi)^{\d/2} E^{(4-\d)} }\right]^{N_\lambda^{(1)} + N_\lambda^{(2)}}\,,
\label{c1}
\end{align}
where $E$ is the center-of-mass energy, and $N^{(1,2)}_{g,y,\lambda}$ are the order in the coupling constants of the two amplitudes. The cross section $\sigma$ has dimension $2-\d$ in $\d$ spacetime dimensions. Eq.~(\ref{c1}) is the NDA  rule --- or master formula --- for cross sections. One can derive this expression by doing the phase space integrals in $\d$ dimensions, or by using the NDA form Eq.~(\ref{nda}) and the optical theorem. As an example, for a generic scalar $\phi$, the $\phi \phi \to \phi \phi $ cross section from a $\lambda \phi^4$ interaction is
\begin{align}
\sigma \sim   \pi (4\pi)^{\d/2} E^{2-\d} \left[\frac{\lambda}{(4 \pi)^{\d/2} E^{(4-\d)} }\right]^{2} \sim  \lambda^2  \frac{\pi}{(4\pi)^{\d/2}}  E^{\d-6},
\label{68}
\end{align}
which gives $\sigma \sim \lambda^2/(16 \pi E^2)$ in $\d=4$.

Note that the final result Eq.~(\ref{c1}) depends on $N_\Lambda$, and not on the number of derivatives in the amplitude. The factors of $4\pi$ for each field, and the different mass dimensions of scalar and fermion fields drop out: the extra $4\pi$ factors for each field cancel the extra $4\pi$ factors in the final particle phase space. Thus for a generic complex scalar $\Phi$, the dimension-six operators in $\d = 4$ with $N_\Lambda=-2$ 
\begin{align}
\cO^{(6)}_{1}&=\frac{(4\pi)^4 \left(\Phi^\dagger \Phi \right)^3}{\Lambda^2} ,& 
\cO^{(6)}_{2}&=\frac{(4\pi)^2 \left(\Phi^\dagger \partial_\mu \Phi \right)^2}{\Lambda^2}
\end{align}
give 
\begin{align}
\sigma  \sim   \pi (4\pi)^2 E^{-2} \left( \frac{E^2}{\Lambda^2}\right)^{2} 
\label{c2}
 \end{align}
 for the $\Phi \Phi \to 4 \Phi$ and $\Phi \Phi \to \Phi \Phi$ cross sections, respectively, where the $\Phi$ mass has been neglected and we have assumed the same coupling strength for the two operators.  Under the same assumptions, dimension eight operators such as 
 \begin{align}
\cO^{(8)}_{1}&=\frac{(4\pi)^6 \left(\Phi^\dagger \Phi \right)^4}{\Lambda^4} , & 
\cO^{(8)}_{2}&=\frac{(4\pi)^2 \left(\Phi^\dagger \partial^2 \Phi \right)^2}{\Lambda^4}
\end{align}
give
 \begin{align}
\sigma  \sim   \pi (4\pi)^2 E^{-2} \left( \frac{E^4}{\Lambda^4}\right)^{2} \,
\label{c23}
 \end{align}
 for the $\Phi \Phi \to 6 \Phi$ and $\Phi \Phi \to \Phi \Phi$ cross sections, respectively.
 
The size of cross sections is thus governed by the usual EFT power counting in $\Lambda$. Dimension six operators give $\sigma \propto 1/\Lambda^4$, dimension eight operators give $\sigma \propto 1/\Lambda^8$, etc. The two operators $\cO^{(6)}_{1,2}$ are both dimension-six operators with $N_\Lambda=-2$, but they have different chiral dimension, $N_\chi=0$ and $N_\chi=2$, respectively, since $N_\chi$ depends on the number of derivatives, Eq.~(\ref{19}). The cross section size is controlled by the $\Lambda$ power counting, \emph{not} by $N_\chi$ (or equivalently $N_p$) counting. Operators with different $N_\chi$ values give the same cross section.

We have shown earlier in Eq.~(\ref{reln2a}) that $R_\Lambda$, $R_\chi$ and $R_F$ are not independent counting rules. The cross section \emph{does not depend} on the number of external fields, as can be seen from Eqs.~(\ref{c2}) and (\ref{c23}).\footnote{Of course, we are only making dimensional arguments here. The actual numerical value can vary with the process.}
It is controlled by the $\Lambda$ power counting. 

Momentum $p$ counting originated in $\chi$PT, and it is instructive to compare $\Lambda$ and $p$ counting in this special case. The chiral field $U=\exp 2 i \vpi/f$ obeys NDA counting. The NDA normalization for a chiral Lagrangian is~\cite{Manohar:1983md}
\begin{align}
\LL &=  f^2 \Lambda^2 \left[ \frac{\partial }{\Lambda} \right]^{N_p} \left[ U \right]^{N_U}
=\frac{\Lambda^4}{16\pi^2} \left[ \frac{\partial }{\Lambda} \right]^{N_p} \left[U \right]^{N_U}
\label{62}
\end{align}
where $N_U$ is the number of powers of $U$ or $U^\dagger$,
and this term has $N_\Lambda=4-N_p$ and $N_\chi=N_p$. Thus the $N_\Lambda$ and $N_\chi$ counting are equivalent --- one can count powers of $\Lambda$ in the denominator or powers of $p$ in the numerator. The $\chi$PT momentum expansion is in powers of $p/\Lambda$, not $p/f$, and so holds up to energies several times $f$.  In deriving this relation, we have treated $U$ as a dimensionless field, or equivalently $\vpi/f \sim \theta$ as a dimensionless angle, which is consistent with NDA. This is often a useful way to think of the chiral field, when one is not interested in a perturbative expansion in powers of the pion field.  For example, in the trace anomaly, one treats $\vpi/f$ as a field with mass dimension zero, rather than $\vpi$ as a field with mass dimension one~\cite{Chivukula:1989ze,Chivukula:1989ds}. Similarly, in studying chiral solitons, one uses $U(x)$ without expanding in $\vpi/f$~\cite{Skyrme:1961vq}. For perturbative calculations, such as pion cross sections, Eq.~(\ref{62}) is expanded in $\vpi/f$.

We now expand out $U$ in powers of $\vpi$, and consider the counting rules for the expanded Lagrangian. The chiral Lagrangian terms we consider are the kinetic and mass terms, and one four-derivative term,
\begin{equation}
\begin{aligned}
\LL &= \LL_k + \LL_m + \LL _4 + \ldots\,,\\
\LL_k &= \frac{f^2}{4} \tr \left( \partial_\mu U \partial^\mu U^\dagger  \right)\,,\\
\LL_m &= \widehat c_m \frac{f^2}{4} \Lambda m_q \left( U  +  U^\dagger \right)\,,\\
\LL_4 &=  \frac{\widehat c_4}{16 \pi^2} \left[ \tr \left( \partial_\mu U \partial^\mu U^\dagger \right) \right]^2\,.
\end{aligned}
\end{equation}
The kinetic term coefficient is fixed so that the pion kinetic term is canonically normalized. The other terms have NDA normalization, so we expect $\widehat c_m$ and $\widehat c_4$ to be order unity. The usual normalization of low energy constants such as $\LL_4$ does not include the $16\pi^2$ of NDA, which is why the coefficients are $\sim 5 \times 10^{-3}$~\cite{Pich:1995bw} for the usual normalization, instead of order unity.  Expanding in powers of $\vpi$ gives schematically
\begin{equation}
\begin{aligned}
\LL_k &\sim \partial^2 \vpi^2 + \frac{\partial^2 \vpi^4}{f^2} +  \frac{\partial^2 \vpi^6}{f^4} + \ldots\,, \\
\LL_m &\sim \widehat c_m \Lambda m_q \left( \vpi^2 + \frac{\vpi^4}{f^2} + \frac{\vpi^6}{f^4}  + \ldots\right)\,,  \\
\LL_4 &\sim  \widehat c_4 \left( \frac{\partial^4 \vpi^4}{\Lambda^2 f^2} +  \frac{\partial^4 \vpi^6}{\Lambda^2 f^4} + \ldots  \right)\,,
\end{aligned}
\end{equation}
using $\Lambda=4\pi f$. $\LL_k$ is a $N_\chi=2$ term and is conventionally referred to as $\mathcal{O}(p^2)$, and $\LL_4$ is $N_\chi=4$ and is called $\mathcal{O}(p^4)$. $\LL_m$ gives a pion mass of order $m_\pi^2 \sim \Lambda m_q$. The term $\LL_k$ with $N_\chi=2$ in expanded form contributes terms  with $N_\Lambda=0,-2,-4,\ldots$ which are of different orders in the $\Lambda$ power counting.

It is conventional to treat $m_q^2$ and $\LL_m$ as $\mathcal{O}(p^2)$, and use a single counting parameter $p$. However, EFTs  have multiple parameters, and combining distinct parameters into one parameter is not always a good idea. The $\xpi\xpi \to \xpi \xpi$ cross section contributions from $\LL_k$ and $\LL_m$ are
\begin{align}
\sigma_k & \sim \frac{ \pi (4\pi)^2}{E^2}  \frac{E^4}{\Lambda^4}, & \sigma_m & \sim  \frac{ \pi (4\pi)^2}{E^2} \frac{m_\pi^4}{\Lambda^4},
\label{65new}
\end{align}
respectively, and have different energy dependence. In the energy regime $m_\pi \ll E \ll \Lambda$, we have $\sigma_k \gg \sigma_m$. The two cross sections are comparable only near threshold, where $E\sim m_\pi$. The systematic counting of powers of symmetry breaking parameters such as $m_q$ is well-known.

Let us return to the momentum expansion. The $\xpi \xpi \to \xpi \xpi$ cross section from $\LL_4 \LL_k$ is
\begin{align}
\sigma_4(\xpi \xpi \to \xpi \xpi) & \sim \frac{ \pi (4\pi)^2}{E^2}  \frac{E^6}{\Lambda^6} \,,
\end{align}
and from $\LL_4 \LL_4$ is
\begin{align}
\sigma_4(\xpi \xpi \to \xpi \xpi) & \sim \frac{ \pi (4\pi)^2}{E^2}  \frac{E^8}{\Lambda^8} \,.
\end{align}
In the $R_\chi$ counting, $\LL_4$ is suppressed by $p^2$ relative to the kinetic Lagrangian $\LL_k$, so $\sigma_4(\xpi \xpi \to \xpi \xpi)$ is suppressed by $E^4/\Lambda^4$ relative to $\sigma_2(\xpi \xpi \to \xpi \xpi)$.  If we instead count $\Lambda$, the $\vpi^4$ term in $\LL_4$ is $1/\Lambda^4$ relative to the $\vpi^4$ term in $\LL_k$, again giving a relative $E^4/\Lambda^4$ suppression. 

Next consider $\xpi \xpi \to 4 \xpi$ from $\LL_k$, 
\begin{align}
\sigma_k(\xpi \xpi \to 4 \xpi ) & \sim \frac{ \pi (4\pi)^2}{E^2}  \frac{E^8}{\Lambda^8} .
\label{67new}
\end{align}
The $6\vpi$ vertex in $\LL_k$ is a $N_\chi=2$ amplitude which is order $p^2$. Nevertheless, the $\LL_k$ contribution to the $\xpi \xpi \to 4 \xpi$ cross section in Eq.~(\ref{67new}) is suppressed by a factor of $E^4/\Lambda^4$ with respect to the $\LL_k$  contribution to the $\xpi \xpi\to \xpi \xpi$ cross section in Eq.~(\ref{65new}). The extra energy suppression arises from the final state phase space. Thus terms with the same $N_\chi$ counting lead to  cross sections of different orders. This result can be explained using the optical theorem.  The cross section $\sigma(\xpi \xpi \to \xpi \xpi)$ is the cut part of a one-loop diagram, whereas $\sigma(\xpi \xpi \to 4 \xpi )$ is the cut part of a three-loop diagram. Since the $R_\chi$ rule depends on the number of loops, the two extra loops add a power $E^4$.

The difference between power counting the amplitude and the cross section is related to  final state phase space, or equivalently, to the number of fields. For a given physical process in $\chi$PT, $N_\chi$ and $N_\Lambda$ counting are equivalent, since the external fields are fixed.  A difference arises only when we compare processes with different numbers of external fields. $N_\chi$ counts the $p$ (or $E$) dependence of the amplitude, and $N_\Lambda$ counts the $p$  (or $E$) dependence of the cross section.

\section{HEFT}\label{sec:heft}

Finally, we consider power counting in HEFT, which describes an extended class of Higgs boson models from the SM to technicolor-like theories and composite Higgs boson models with a light scalar $h$.  The SMEFT is a special case of HEFT when the scalar manifold has an $O(4)$ (or $SU(2) \times U(1)$ if we do not assume custodial symmetry) invariant fixed point~\cite{Alonso:2015fsp}.

HEFT extends the chiral Lagrangian of the three ``eaten" Goldstone bosons of the SM gauge theory~\cite{Appelquist:1980vg,Longhitano:1980iz,Longhitano:1980tm,Feruglio:1992wf} by including a light physical scalar singlet $h$.   The three Goldstone bosons appear in the HEFT Lagrangian through $\U(x) \equiv \exp2 i \vpiEW(x)/f$, where $\vpiEW(x)$ is the Goldstone boson matrix and $f$ is the Goldstone boson decay constant.  Inspired by composite Higgs models in which the $h$ field is also a Goldstone boson, the dependence of operators on $h/f$ is customarily encoded by a generic polynomial function $\cF(h)$~\cite{Grinstein:2007iv}, which has a power series expansion in $h/f$. It also is customary to define the vector and scalar chiral fields
\begin{equation}
\V_\mu\equiv \left(D_\mu \U\right)\U^\dag\,,\qquad\qquad
\T\equiv \U\sigma_3\U^\dag\,,
\end{equation}
which transform as the adjoint representation of the $SU(2)_L$ gauge symmetry and are custodial $SU(2)$ preserving and breaking, respectively.

The field $\U(x)$ has an expansion in powers of $\vpiEW(x)/f$, so it produces operators with different mass dimension. As we have shown in Sec.~\ref{sec:compare}, these operators contribute to cross sections at \emph{different} orders in $E/\Lambda$, so  one cannot treat $\U$ as having a homogeneous mass dimension for power counting purposes. The expansions of $\U$, $\V$ and $\T$ are
\begin{equation}
\begin{aligned}
\U &= \openone + \frac{2i}{f}  \vpiEW + \ldots \\
\V_\mu &= \frac{2i}{f} \partial_\mu \vpiEW +  \frac{2i}{f} \left[\vpiEW,g A_\mu\right] + \frac{gv}{f} B_\mu+\ldots  \\
\T &= \sigma_3 +   \frac{2i}{f} \left[ \vpiEW ,\sigma_3 \right] +\ldots
\end{aligned}
\label{621}
\end{equation}
where $A_\mu$ are the unbroken gauge fields, and $B_\mu$ are the broken gauge fields, and we have retained the lowest dimension non-vanishing terms in each expansion. We will define the primary dimension $d_p$ of each HEFT operator as the smallest  operator mass dimension of the terms resulting from its power series expansion. With this definition $\U$ and $\T$ have primary dimension zero and $\V_\mu$ has primary dimension two. 

The rule for $\V_\mu$ is a bit subtle. The operators $\partial_\mu \vpiEW$ and  $ \left[\vpiEW,g A_\mu\right]$ have mass dimension two, while the broken gauge field $B_\mu$ has mass dimension one.  With generic power counting, the broken gauge boson masses are of order $f$. However, HEFT has a vacuum alignment fine-tuning, so the broken gauge boson masses are of order $v$ instead of $f$. Thus the broken gauge field term $gB_\mu$ is multiplied by $v/f$. Finally, all the leading terms stemming from $\V_\mu$ have mass dimension 2 divided by the scale $f$. Another way to say this is that the longitudinal components of broken gauge bosons behave like $\partial \vpiEW/f$, as can be seen from the $k_\mu k_\nu/M^2$ term in the propagator.  It is precisely this longitudinal component, related to $\partial \vpiEW/f$ by the Equivalence Theorem~\cite{Cornwall:1974km,Vayonakis:1976vz,Lee:1977eg,Chanowitz:1985hj}, which appears in $\V_\mu$. The counting of the $B_\mu$ field strength tensor is not affected, since it only depends on the transverse part. The $\V_\mu$ term in Eq.~(\ref{621}) has both broken and unbroken gauge bosons since the HEFT formalism uses the analog of the $\Sigma$-basis for QCD $\chi$PT.  A more elegant formalism using the $\xi$-basis~\cite{Coleman:1969sm,Callan:1969sn,Georgi:1985kw} splits $\V_\mu$ into $\mathscr{V}_\mu$, which is part of the chiral covariant derivative and transforms as a gauge field under the broken symmetry, and $\mathscr{A}_\mu$, which transforms as an adjoint under the broken symmetry. The  chiral covariant derivative $\mathscr{D}_\mu = \partial_\mu + \mathscr{V}_\mu$ has dimension one, and $\mathscr{A}_\mu$ has dimension two.

Following Refs.~\cite{Azatov:2012bz, Alonso:2012px, Alonso:2012pz,  Gavela:2014vra, Buchalla:2013rka}, examples of  HEFT lepton and baryon number preserving operators are shown schematically in Table~\ref{tab:opsChiral}, using NDA normalization in $\d=4$ spacetime dimensions and ordering the operators by increasing primary dimension $d_p$.  Also shown are the values of $N_\chi$. The Higgs functions $\cF_i(h)$ are treated as dimensionless functions analogous to $\U$ and are normalized so that $\cF_i(0)  = 1$. The primary dimension $d_p$ of $\partial \cF(h)$ is $d_p=2$, since the expansion starts with $\partial h/f$. Operators in the $\xi$-basis are given by using $\mathscr{D}$ for the derivatives, $\V_\mu \to \mathscr{A}_\mu$ and $\U \to 1$.

The importance of operators cannot be determined by the explicit powers of $1/\Lambda$ in front of the operator in Table~\ref{tab:opsChiral}, because there are hidden factors of $\Lambda$ in $\U$, $\V$ and $\T$, which become manifest when they are expanded in $\Pi$ as in Eq.~(\ref{621}).  In the large $\Lambda$ limit, all terms can be expanded out in a power series in $\vpiEW/f=4 \pi \vpiEW/\Lambda$, and the HEFT Lagrangian reduces to SMEFT form, where the $1/\Lambda$ counting is manifest. For example, the $d_p=8$  term $\V^4$, which has no explicit power of $1/\Lambda$ in Table~\ref{tab:opsChiral}, is $(\partial \vpiEW)^4/f^4$ at leading order in its power series expansion, which is suppressed by $1/\Lambda^4$, as expected for a $d=8$ operator.

\begin{table}
\centering
\renewcommand{\arraystretch}{1.5}
\begin{tabular}{c|c|c|c}
Operator& $d_p$ &  $N_\chi$ & NDA form\\
\hline
\hline
$\openone$
& $0$ 
& $0$
& $\frac{\Lambda^4}{(4\pi)^2} \, \cF_{\openone}(h)$
\\
$\psi^2 \U$
&$3$ 
& $1$
&$\Lambda \, \psi^2 \U \, \cF_{\psi^2 \U}(h)$
\\
$X^2$ 
& $4$  
& $2$
&$X^2 \, \cF_{X^2}(h)$
\\
$\psi^2 D$ 
& $4$ 
& $2$
&$\psi^2 D$
\\
$( \partial h)^2$
&$4$ 
&$2$
&$(\partial h)^2$
\\
$\V^2 $
&$4$ 
&$2$
&$\frac{\Lambda^2}{(4\pi)^2} \, \V^2 \, \cF_{\V^2}(h)$
\\
\hline
$\psi^2 \V $
&$5$ 
&$2$
&$\psi^2 \V \, \cF_{\psi^2 \V}(h)$
\\
$\psi^2 X \U$
&$5$
&$2$
&$\frac{4\pi}{\Lambda}\, \psi^2 X \U \, \cF_{\psi^2 X \U}(h)$
\\
\hline
$\psi^4$
&$6$ 
&$2$
&$\frac{(4\pi)^2}{\Lambda^2} \, \psi^4  \, \cF_{\psi^4}(h)$
\\
$X \V^2$
&$6$ 
&$3$
&$\frac{1}{4\pi} \, X \V^2 \, \cF_{X \V^2}(h)$
\\
$X^3$
&$6$ 
&$3$
&$\frac{(4\pi)}{\Lambda^2} \, X^3 \, \cF_{X^3}(h)$
\\
$X \V \partial $
&$6$ 
&$3$
&$\frac{1}{4\pi} \, X \V \, \partial  \cF_{X \V \partial }(h)$
\\
\hline
$\psi^2 \V \U \partial $
& $7$ 
&$3$
&$\frac{1}{\Lambda} \, \psi^2 \V \U  \, \partial \cF_{\psi^2 \V \U \partial }(h)$
\\
$\psi^2 \V^2 \U $
&$7$ 
&$3$
&$\frac{1}{\Lambda} \, \psi^2 \V^2 \U \, \cF_{\psi^2 \V^2 \U}(h)$
\\
$\psi^2 \U \partial^2 $
& $7$ 
&$3$
&$\frac{1}{\Lambda} \, \psi^2 \U \,  \left(\partial\,  \cF_{\psi^2 \U \partial^2 }(h)\right)^2$ \\
\hline
$\V^2 \partial^2 $
&$8$ 
&$4$
&$\frac{1}{(4\pi)^2} \, \V^2 \, \left(\partial \cF_{\V^2 \partial^2 }(h)\right)^2$  \\
$\V^4$
&$8$ 
&$4$
&$\frac{1}{(4\pi)^2} \, \V^4 \, \cF_{\V^4}(h)$
\\
\end{tabular}
\caption{\it   Custodial-preserving $CP$-even HEFT operators of primary dimension $d_p$ and chiral number $N_\chi \equiv N_p + N_\psi/2$, normalized using NDA rule Eq.~(\ref{nda4}) in $\d=4$ spacetime dimensions.  The notation is schematic, with $h$ the physical Higgs singlet field, $\psi$ a fermion field, $X_{\mu \nu}$ a field-strength tensor, $D$ a covariant derivative,  $\partial$ a partial derivative, $\U(x)$ the exponential of the Goldstone boson matrix, and $\V(x)$ the vector chiral field. All indices are suppressed. The operators listed have $d_p \le 8$ and $N_\chi \le 4$.
\label{tab:opsChiral}}
\end{table}

The custodial-preserving Lagrangian of HEFT including terms with $d_p \le 4$ is
\begin{align}
\LL^{(d_p \le 4)}=& -\frac14 X^2 \, \cF_{X^2}(h)+ \overline \psi i\slashed{D} \psi + \frac 12 \partial_\mu h \partial^\mu h\\
&-\frac{f^2}{4} \, \text{Tr}(\V_\mu\V^\mu) \, \cF_{\V^2}(h) 
+\, \widehat C_{\openone} \,  {f^2\Lambda^2} \, \cF_{\openone}(h)\nonumber\\
& -\,\widehat C_{\psi^2 \U}\, \Lambda  \left(\bar\psi_L\U\psi_R+\text{h.c.}\right) \, \cF_{\psi^2 \U}(h)\,, \nonumber
\end{align}
where we have set $\Lambda=4\pi f$ and the term proportional to $\widehat C_{\openone}$ encodes the Higgs scalar potential. There are no arbitrary $\cF(h)$ functions in front of the fermion and Higgs kinetic energies because they can be removed by field transformations (see Refs.~\cite{Brivio:2013pma,Brivio:2014pfa,Gavela:2014vra}).  In addition, there is no arbitrary coefficient in front of the $\V^2$ term, since the coefficient must be unity to produce a canonically normalized kinetic energy term for the $\vpiEW$ fields. The $\psi^2 \U$ term is a chirality violating operator which gives mass to the fermions, so its coefficient defines a Yukawa coupling
\begin{align}
\widehat C_{\psi^2 \U} &= \hat y \equiv \frac{y}{4\pi} .
\end{align}
Note that
\begin{eqnarray}
\widehat C_{\psi^2 \U} \Lambda &=& \hat y \Lambda \equiv y f 
\end{eqnarray}
which converts  $\Lambda$ to $f$ if the operator is written in terms of the standard Yukawa coupling $y$ instead of the NDA Yukawa coupling $\hat y$. As for the gauge boson masses, a fine-tuning $v/f$ is required in the Yukawa interactions in order to predict fermion masses proportional to the EW vev $v$.

The HEFT counting discussed in this section also can be used for QCD $\chi$PT which corresponds to setting all $\cF_i(h)=1$. The leading order (LO) Lagrangian contains the two-derivative pion terms and photon kinetic energy term. 
The terms
\begin{align}
&H_1 \left[ \tr F_{R\mu \nu}^2 + \tr F_{L\mu \nu}^2\right], &
& L_{10}  \tr \U^\dagger F_{R\mu \nu} \U F_{L\mu \nu}\,,
\end{align}
(in the notation of Ref.~\cite{Pich:1995bw}) give the the running of the photon kinetic energy due to pion loops, and naturally belong with the photon kinetic energy term in the $d_p=4$ Lagrangian. The two-derivative plus one field-strength term $L_{9}$ is in the $d_p=6$ Lagrangian, while the other chiral symmetry preserving operators $L_{1-3}$ describing four-derivative pion interactions are contained in the $d_p=8$ one. If the quark mass term is treated as order $p^2$, then the chiral symmetry breaking operators $H_2$ and $L_{6-8}$ are in the $d_p=4$ Lagrangian, while the terms $L_{4,5}$ are in the $d_p=6$ one.

\subsection{Loops in HEFT}

The SMEFT Lagrangian can be broken up into a leading order Lagrangian $\LL_{\text{LO}}$ with terms of $d \le 4$, a NLO Lagrangian with operators of $d=6$, a NNLO Lagrangian with operators of $d=8$, etc. The $\Lambda$ power counting implies that loops with $\LL_{\text{LO}}$ vertices generate divergent contributions only to LO operators, loops with one insertion of $\LL_{\text{NLO}}$ generate divergent contributions only to NLO operators, loops with two insertions of $\LL_{\text{NLO}}$ or one insertion of $\LL_{\text{NNLO}}$ generate divergent contributions only to NNLO terms, etc. Note that the LO, NLO, etc.\ counting does not depend on the number of loops in the diagram. Thus an arbitrary loop graph using $\LL_{\text{LO}}$ vertices only generates $\LL_{\text{LO}}$ operators.

In $\chi$PT, the Lagrangian can be broken up into the $N_\chi=2$ leading order Lagrangian,
\begin{align}
\LL &= \frac{f^2}{4} \tr\, \partial_\mu \U \, \partial^\mu \U^\dagger\,,
\end{align}
the $N_\chi=4$ order $p^4$ NLO Lagrangian, etc. $\partial_\mu \U(x)$ contains terms with different mass dimension, all with one derivative. The $N_\chi$ counting rule Eq.~(\ref{chi}) implies that a graph with arbitrary insertions of $\LL_{\text{LO}}$ vertices, \emph{but only one loop} generates divergent contributions to $\LL_{\text{NLO}}$ terms; graphs with arbitrary $\LL_{\text{LO}}$ vertices, one insertion of $\LL_{\text{NLO}}$ plus one loop, or arbitrary $\LL_{\text{LO}}$ vertices and two loops, generates divergent contributions to $\LL_{\text{NNLO}}$ terms; etc.

The usual SMEFT and $\chi$PT expansions are both systematic, but different. The SMEFT power counting does not depend on the number of loops, but does depend on the number of fields, whereas the $\chi$PT power counting depends on the number of loops but not on the number of fields. The breakup of the two Lagrangians into LO, NLO, etc.\ is also different; one counts fields, and the other counts derivatives. The $N_\Lambda$ and $N_\chi$ counting rules are both equally valid in both theories; what differs is the way terms are grouped together.  In $\chi$PT, the $N_\chi$ rule is more convenient because all terms in $\U(x)$ have the same $N_\chi$ value.

HEFT is a fusion of  SMEFT and $\chi$PT. It contains Goldstone boson fields in $\U(x)$ as well as other fields such as gauge fields, fermions, and the Higgs boson $h$, which makes  a unified power counting more subtle. To see the problem, consider a toy theory which is QCD with charge zero quarks, and QED with the muon integrated out so that there are higher dimension operators suppressed by inverse powers of the muon mass. At low energies, the QCD sector is described by $\chi$PT with $N_\chi$ counting and an expansion in $p/\Lambda$. The lepton sector is QED with \emph{two} expansion parameters, $\alpha$ and $p/m_\mu$. One-loop graphs with the  $\mathcal{O}(p^2)$ LO $\chi$PT Lagrangian generate $\mathcal{O}(p^4)$ terms of the  NLO $\chi$PT Lagrangian. One loop graphs with the  $\mathcal{O}(1/m_\mu^0)$ LO QED Lagrangian generate $\mathcal{O}(1/m_\mu^0)$ terms of the LO QED Lagrangian with coefficients suppressed by $\alpha$, not $\mathcal{O}(1/m_\mu^2)$ terms of the NLO QED Lagrangian. The two sectors satisfy two different types of power counting rules with different expansion parameters, one based on QED counting, and the other based on $\chi$PT counting. Turning on an electric charge for the quarks couples the two sectors, but it is not helpful to force both sectors into a unified counting with a single expansion parameter. The situation in HEFT is similar.

The primary dimension used to classify the operators uses a SMEFT-like counting of dimension combined with summing up powers of $\vpiEW/f$ into $\U(x)$ to maintain the symmetry transformation properties of the chiral field. Consider the $\V^2$ operator in Table~\ref{tab:opsChiral}, with expansion
\begin{equation}
\begin{split}
f^2 \, \text{Tr}(\V_\mu\V^\mu) \, \cF_{\V^2}(h) \sim& (\partial \vpiEW)^2 +c_1 \frac{1}{f} h  (\partial \vpiEW)^2 +\\
&+  c_2 \frac{1}{f^2} (\partial \vpiEW)^2 ( \vpiEW)^2 + \ldots
\end{split}
\end{equation}
which has $d_p=4$, and contains operators with dimensions $d=d_p + k$, $k \ge 0$, although all terms have $N_\chi=2$. Using chiral operators such as $\V^2$ in graphs produces an operator which must be written in terms of $\U$ by chiral invariance, with
\begin{equation}
\begin{aligned}
d_p - 4 & \ge \sum_i (d_{p,i}-4) \\
N_\chi - 2 &= \sum_i ( N_{\chi,i}-2) + 2 L
\end{aligned}
\label{two}
\end{equation}
from Eqs.~(\ref{Lambda}) and (\ref{chi}). The $\Lambda$ counting rule becomes an inequality when written in terms of $d_p$ instead of $d$. 

The primary dimension is a way of ordering terms in the Lagrangian for phenomenological applications, while at the same time treating objects such as $\V_\mu(x)$ with terms related by chiral invariance as a single quantity. The underlying counting rules remain the four independent rules summarized in Sec.~\ref{Sect:MasterFormula}.

\section{Conclusions}\label{sec:conclusions}

We have derived the most general power counting rules for EFT, and shown how to use them in a number of examples. We have clarified the difference between $\Lambda$ and chiral number $N_\chi$ counting and shown that cross sections are controlled by the $\Lambda$ counting, not the $N_\chi$ counting. We have applied the rules to HEFT, and clarified some aspects of HEFT and $\chi$PT power counting. The ordering of cross sections in HEFT is by the $\Lambda$ power counting, and hence by the primary dimension $d_p$ listed in Table~\ref{tab:opsChiral}. We have also shown the NDA counting is related to $\hbar$ counting.  A generalization of the $\d=4$ relation $\Lambda=4\pi f$ to arbitrary dimensions also has been derived.

\bigskip

\acknowledgments

We  thank I.~Brivio, M.J.~Herrero, C.~Murphy, S.~Rigolin, S.~Saa and G.~Zanderighi for useful discussions. This work was supported in part by grants from the Simons Foundation (\#340282 to Elizabeth Jenkins and \#340281 to Aneesh Manohar), Spanish MINECO's ``Centro de Excelencia Severo Ochoa'' Programme under grant SEV-2012-0249, and by DOE grant DE-SC0009919. M. Belen Gavela and Luca Merlo acknowledge partial support of the European Union network FP7 ITN INVISIBLES (PITN-GA-2011-289442),
FP10 ITN ELUSIVES (H2020-MSCA-ITN-2015-674896) and INVISIBLES-PLUS (H2020- MSCA-RISE-2015-690575), and of CiCYT through 
the project FPA2012-31880.


%

\end{document}